\documentclass[aps,pre,showpacs,noshowkeys,amsmath,amssymb,amsfonts,superscriptaddress,longbibliography,reprint]{revtex4-1}
\usepackage[english]{babel}

\usepackage{graphicx}
\usepackage{bm}
\usepackage{physics}
\usepackage{mathtools}
\usepackage{gensymb}

\setcitestyle{super}
\usepackage{caption}
\usepackage{subcaption}
\DeclareCaptionLabelSeparator{bar}{~\rule[-0.4ex]{0.2ex}{1em}~}
\DeclareCaptionLabelFormat{subfor}{\textbf{#2}}
\newcommand*\bfcaption[2]{\caption[#1]{\textbf{#1.}#2}}
\usepackage{xcolor}
\definecolor{UBcolor}{HTML}{007CC1}
\newcommand{\MYhref}[2]{\href{#1}{\color{UBcolor}{#2}}}
\usepackage[colorlinks=true,pdfnewwindow=true,linkcolor=UBcolor,citecolor=UBcolor,urlcolor=UBcolor,breaklinks=true,linktocpage]{hyperref}
\usepackage[all]{hypcap}
\usepackage[nameinlink,capitalise]{cleveref}
\usepackage{xr}
\crefname{SI section}{SI Section}{SI Sections}
\Crefname{SI section}{SI Section}{SI Sections}
\begin{document}

\title{Flocking by turning away}

\author{Suchismita Das}
\affiliation{Max Planck Institute for the Physics of Complex Systems, N\"{o}thnitzerst. 38, 01187 Dresden, Germany}

\author{Matteo Ciarchi}
\affiliation{Max Planck Institute for the Physics of Complex Systems, N\"{o}thnitzerst. 38, 01187 Dresden, Germany}

\author{Ziqi Zhou}
\affiliation{Department of Polymer Science and Engineering, University of Science and Technology of China (USTC), Hefei, Anhui, 230026, China}

\author{Jing Yan}
\affiliation{Department of Molecular, Cellular and Developmental Biology, Yale University, New Haven, CT, USA}
\affiliation{Quantitative Biology Institute, Yale University, New Haven, CT, USA}

\author{Jie Zhang}
\email{zhjie717@ustc.edu.cn}
\affiliation{Key Laboratory of Precision and Intelligent Chemistry, University of Science and Technology of China (USTC), Hefei, Anhui, 230026, China}
\affiliation{Department of Polymer Science and Engineering, University of Science and Technology of China (USTC), Hefei, Anhui, 230026, China}

\author{Ricard Alert}
\email{ralert@pks.mpg.de}
\affiliation{Max Planck Institute for the Physics of Complex Systems, N\"{o}thnitzerst. 38, 01187 Dresden, Germany}
\affiliation{Center for Systems Biology Dresden, Pfotenhauerst. 108, 01307 Dresden, Germany}
\affiliation{Cluster of Excellence Physics of Life, TU Dresden, 01062 Dresden, Germany}

\date{\today}

\begin{abstract}
Flocking, as paradigmatically exemplified by birds, is the coherent collective motion of active agents. As originally conceived, flocking emerges through alignment interactions between the agents. Here, we report that flocking can also emerge through interactions that turn agents away from each other. Combining simulations, kinetic theory, and experiments, we demonstrate this mechanism of flocking in self-propelled Janus colloids with stronger repulsion on the front than on the rear. The polar state is stable because particles achieve a compromise between turning away from left and right neighbors. Unlike for alignment interactions, the emergence of polar order from turn-away interactions requires particle repulsion. At high concentration, repulsion produces flocking Wigner crystals. Whereas repulsion often leads to motility-induced phase separation of active particles, here it combines with turn-away torques to produce flocking. Therefore, our findings bridge the classes of aligning and non-aligning active matter. Our results could help to reconcile the observations that cells can flock despite turning away from each other via contact inhibition of locomotion. Overall, our work shows that flocking is a very robust phenomenon that arises even when the orientational interactions would seem to prevent it.
\end{abstract}

\maketitle

Flocking --- the self-organized collective motion of active agents --- is ubiquituous in Nature \cite{Vicsek2012}. It takes place in many systems across scales, from bird flocks \cite{Cavagna2014} to bacterial colonies \cite{Peruani2012} and to cytoskeletal filaments driven by molecular motors \cite{Schaller2010}. Understood as the emergence of polar order in systems of self-propelled particles, flocking is a landmark phenomenon that launched the field of active matter \cite{Vicsek1995,Toner1995}. As originally conceived in the Vicsek model \cite{Vicsek1995}, flocking arises through alignment interactions between the active agents, which align similarly to spins in the XY model. Alignment-based flocking has been experimentally realized using synthetic active colloids, which feature alignment interactions of either hydrodynamic, electric, or magnetic origin \cite{Bricard2013,Kaiser2017,Geyer2018}.

However, recent work showed that flocking can also emerge without explicit alignment interactions \cite{Shaebani2020,Bar2020,Chate2020,Baconnier2024}. Instead of aligning with neighbors, the agents can experience a variety of alternative interactions \cite{Szabo2006,Ferrante2013,Grossman2008,Deseigne2010,Romanczuk2009,Grossmann2013,Strombom2011,Barberis2016,Barberis2019,Grossmann2020,Knezevic2022,Chen2023a,Casiulis2022,Caprini2023,Kursten2023a,Kopp2023}, such as aligning with their own velocity or force \cite{Szabo2006,Ferrante2013,Baconnier2024}, colliding inelastically \cite{Grossman2008,Deseigne2010}, or chasing others in their vision cone \cite{Strombom2011,Barberis2016,Barberis2019}.

Such alternative interactions were inferred in schooling fish \cite{Katz2011,Jhawar2020}, and they might be more widespread than standard alignment interactions. For example, robots in a swarm might benefit from collision-avoidance interactions that reorient them away from collisions \cite{Hayes2002,Chen2023a}. Similarly, several types of motile cells undergo contact inhibition of locomotion --- a behavior whereby cells repolarize away from cell-cell collisions \cite{Mayor2010,Stramer2017,Alert2020}. Yet, cell layers and trains have been observed to flock, both in simulations \cite{Smeets2016,Hiraiwa2020} and in experiments \cite{Szabo2006,Desai2013,Cetera2014,Malinverno2017,Jain2020c,Brandstatter2023,Lang2024,Tan2022b}. How do cells flock despite interacting via contact inhibition of locomotion? More generally, what types of orientational interactions lead to flocking?

Here, we show that agents that turn away from each other can collectively align and flock. This finding is suprising because turn-away interactions are intuitively expected to prevent and destroy orientational order. We show that this mechanism of flocking requires the combination of turn-away torques and repulsive forces between the particles. Therefore, our findings bridge the classes of alignment-based and repulsion-based phenomena in active matter, respectively represented by flocking \cite{Marchetti2013} and motility-induced phase separation (MIPS) \cite{Cates2015}. Our results expand the types of interactions that can produce flocking, and they might help to understand the physical origin of flocking in cell collectives. More generally, our results demonstrate the emergence of macroscopic polar order from microscopic interactions that do not implement polar alignment. Therefore, our findings strikingly showcase the disconnect between the symmetries of microscopic interactions and macroscopic order in active matter \cite{Huber2018,Grossmann2020,Zhang2021i}.

\bigskip

\noindent\textbf{Flocking of metal-dielectric Janus colloids}

\begin{figure*}[tb]
\begin{center}
\includegraphics[width=\textwidth]{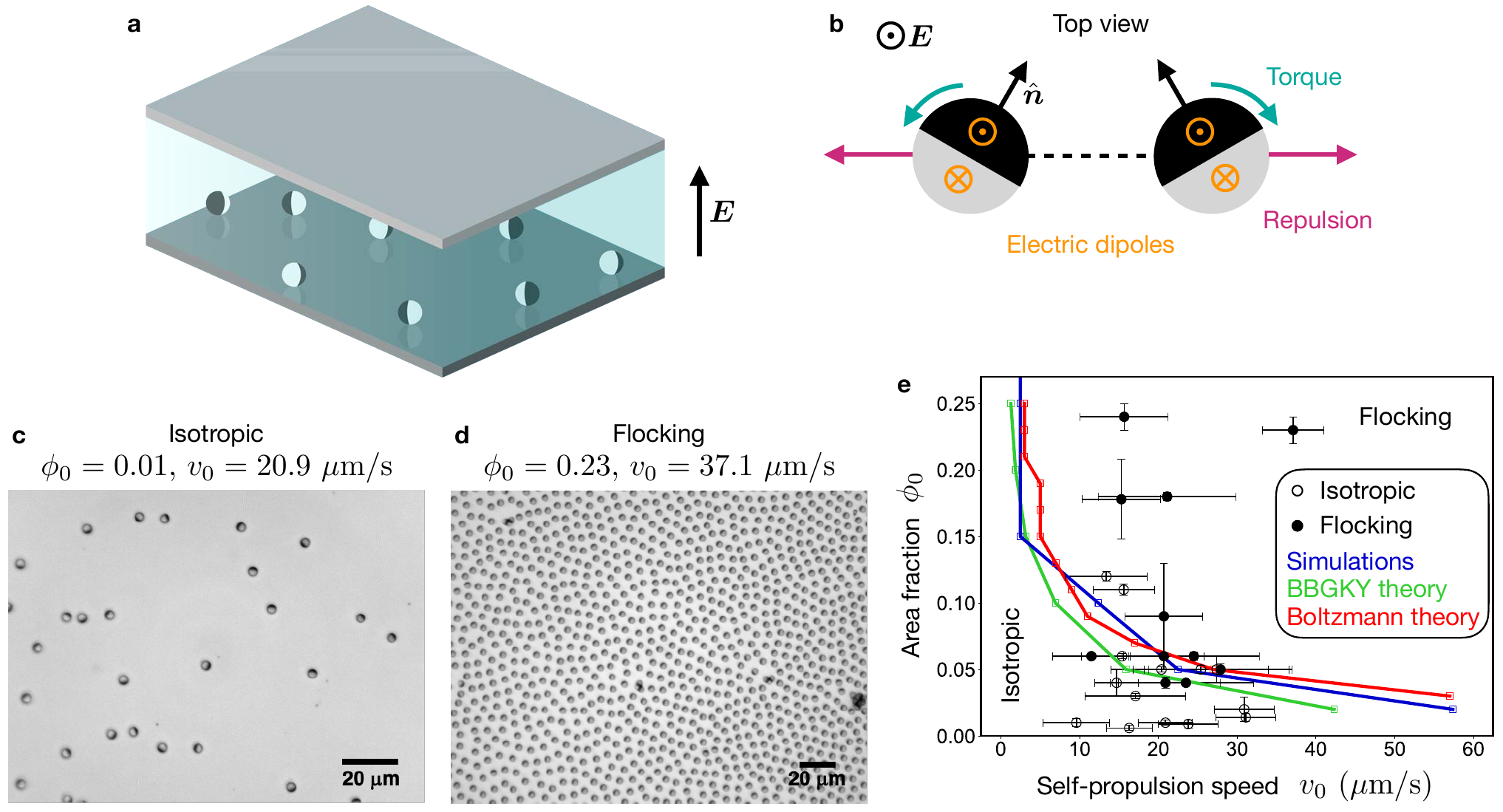}
\end{center}
  {\phantomsubcaption\label{Fig system}}
  {\phantomsubcaption\label{Fig interactions}}
  {\phantomsubcaption\label{Fig isotropic}}
  {\phantomsubcaption\label{Fig flocking}}
  {\phantomsubcaption\label{Fig experimental-diagram}}
\bfcaption{Flocking of metal-dielectric Janus colloids}{ \subref*{Fig system}, Schematic of the experimental setup in which $3$ $\mu$m-diameter particles are allowed to sediment in water to the bottom of a sample cell across which AC electric fields are applied vertically. \subref*{Fig interactions}, Top view of two Janus particles in an electric field $\bm{E}$ that induces dipoles of opposite orientation and different magnitude (orange) on the head and tail hemispheres. This leads to particle self-propulsion along the direction $\hat{\bm{n}}$ (black), and to interparticle forces (purple) and torques (green). Torques rotate particles away from the direction of the interparticle distance (dashed line). \subref*{Fig isotropic},\subref*{Fig flocking}, The system forms an isotropic gas at low area fraction and self-propulsion speed (\subref*{Fig isotropic}), and it flocks at high area fraction and speed (\subref*{Fig flocking}). \subref*{Fig experimental-diagram}, Phase diagram of the flocking transition. Points show the experimental data. The lines indicate the phase boundaries that we predict via our simulations and two theory approaches. The experimental points are averages over time for durations ranging from 25 s to 656 s, at either 10 or 20 frames/second, including between 317 and 13115 frames. Error bars are s.d.} \label{Fig 1}
\end{figure*}

We study a suspension of self-propelled Janus colloidal particles \cite{Yan2016,VanderLinden2019,Zhang2021i,Iwasawa2021}. The particles are $3$ $\mu$m-diameter silica spheres, coated with $35$ nm of titanium and $20$ nm of silicon oxide on one hemisphere (\cref{methods}). These particles are suspended in deionized water and placed between conductive coverslips coated with indium tin oxide, separated by a $120$ $\mu$m spacer (\cref{Fig system}, \cref{methods}). Particles sediment to form a monolayer on the bottom coverslip. To drive the particles, we apply a perpendicular AC voltage of amplitude $V_0=10$ V and frequency $\nu=30$ kHz. The resulting electric field aligns the particle equator perpendicular to the coverslips, and it polarizes the metal and dielectric hemispheres differently (\cref{Fig interactions}). This difference induces (i) electrokinetic flows that produce particle self-propulsion along a direction $\hat{\bm{n}}$ pointing from the dielectric to the metallic hemisphere  \cite{Gangwal2008,Moran2017,Bishop2023}, and (ii) electrostatic interparticle forces and torques (\cref{Fig interactions}).

Upon application of the electric field, the system remains as an isotropic active gas at low area fractions and self-propulsion speeds (\cref{Fig isotropic}, \hyperref[movies]{Movie 1}). In contrast, at higher area fractions and speeds, the system develops polar order and flocks (\cref{Fig flocking}, \hyperref[movies]{Movie 2}). In this regime, we observe spatiotemporal patterns including vortices and large-scale polar bands characteristic of flocking systems (\hyperref[movies]{Movies 3 and 4}).

We analyze different experimental realizations recorded at either high or low magnification. In high-magnification movies, we can track particle orientations $\hat{\bm{n}}_i(t)$ and measure the time-averaged polar order parameter $P = \frac{1}{N} \langle | \sum_{i=1}^N {\hat{\bm{n}}_i}(t) | \rangle_t$ (\cref{Fig experiment-polar-order,Fig velocity-polarity-correlation}). In low-magnification movies, we cannot resolve single-particle orientations, and we instead perform particle image velocimetry (PIV) to measure the flow field $\bm{v}(\bm{r})$ and obtain its correlation length (\cref{Fig correlation-length,Fig correlation-length-analysis}). Based on these measurements, we classify the state of each realization as either isotropic or flocking (\cref{methods}). \Cref{Fig experimental-diagram} shows that our experimental results are consistent with the simulations and theories that we develop below.

\bigskip

\noindent\textbf{Active Brownian particles with turn-away interactions}

\begin{figure}[tb]
\begin{center}
\includegraphics[width=\columnwidth]{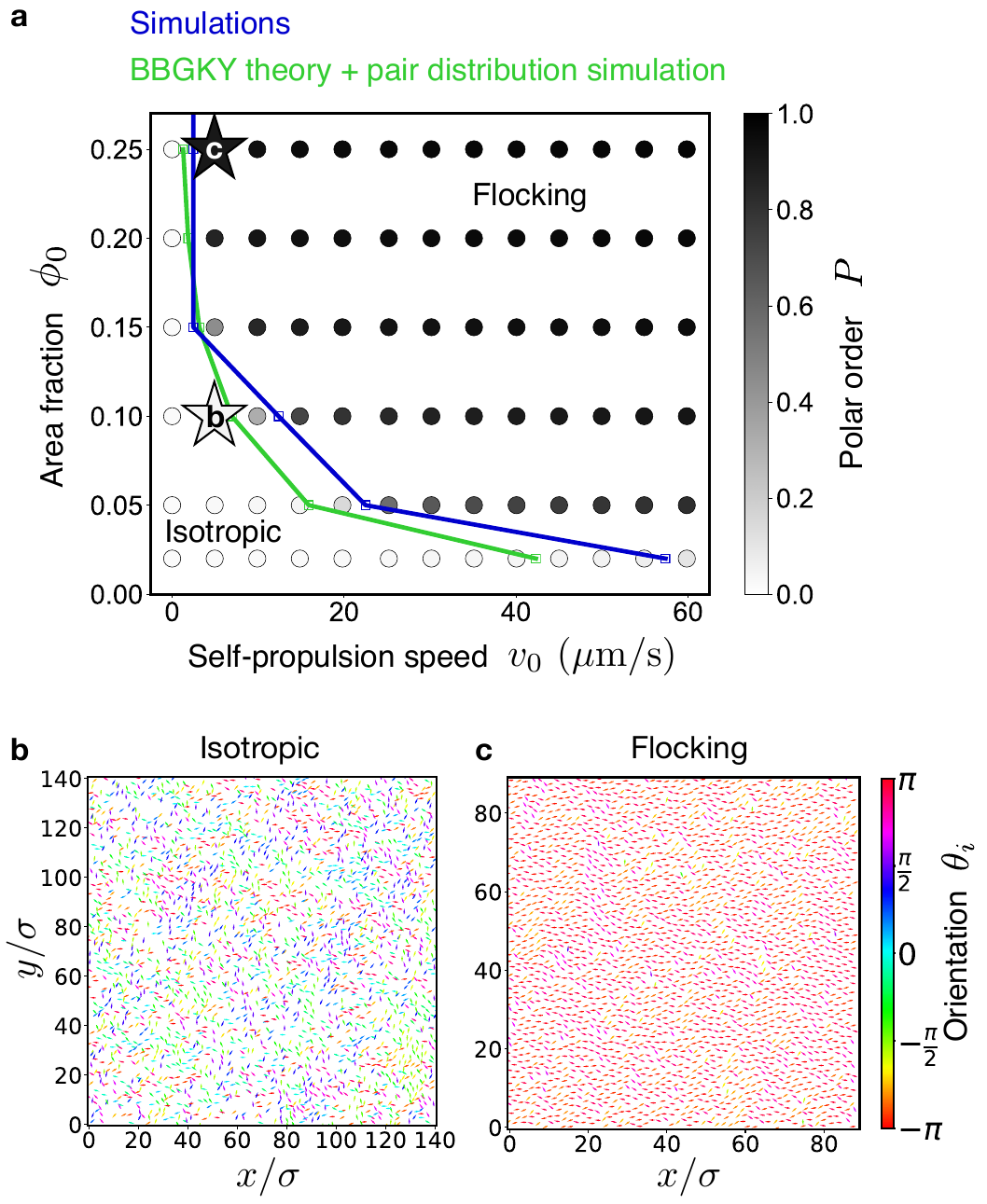}
\end{center}
  {\phantomsubcaption\label{Fig phase-diagram}}
  {\phantomsubcaption\label{Fig isotropic-simulation}}
  {\phantomsubcaption\label{Fig flocking-simulation}}
\bfcaption{Flocking in simulations of repulsive active Brownian particles with turn-away torques}{ \subref*{Fig phase-diagram}, Phase diagram showing the flocking transition as either the area fraction or the self-propulsion speed increases. The blue phase boundary is obtained from the point of steepest ascent of the measured polar order parameter $P$ (\cref{Fig flocking-transition}). The green phase boundary is predicted using BBGKY kinetic theory using the pair distribution function measured in simulations (\crefrange{Fig quadrant-1}{Fig quadrant-4}). Stars indicate the snapshots shown below. \subref*{Fig isotropic-simulation},\subref*{Fig flocking-simulation}, Snapshots showing the isotropic phase (\subref*{Fig isotropic-simulation}) and the polar flocking phase (\subref*{Fig flocking-simulation}). The evolution towards the flocking state is shown in \cref{Fig emergence-flocking} and \hyperref[movies]{Movie 6}.} \label{Fig 2}
\end{figure}

The observation of flocking is surprising as the electrostatic interactions between the particles tend to repel them and turn them away from each other (\cref{Fig interactions}). To investigate if and how these interactions give rise to flocking, we used a two-dimensional microscopic model based on the dipolar interactions between the hemispheres of our particles \cite{Zhang2021i}. Our model shows that two particles interact via a repulsive force
\begin{equation} \label{eq force}
\bm{F}_{ij} = \frac{3}{4\pi\epsilon} \frac{(d_{\text{h}} + d_{\text{t}})^2}{r_{ij}^4} e^{-r_{ij}/\lambda}\; \hat{\bm{r}}_{ij},
\end{equation}
where $\epsilon$ is the dielectric permittivity of the solvent, $\bm{r}_{ij} = \bm{r}_j - \bm{r}_i$ is the distance vector, and $d_{\text{h}}>0$ and $d_{\text{t}}<0$ are the effective dipole strengths of the head and tail hemispheres, respectively (\cref{Fig interactions}). The exponential factor accounts for screening by the electrodes, separated by a distance $\lambda = 120$ $\mu$m. Moreover, because head dipoles are stronger than tail dipoles ($d_{\text{h}}^2> d_{\text{t}}^2$), particles interact via a torque
\begin{equation} \label{eq torque}
\bm{\Gamma}_{ij} = \frac{3\ell}{4\pi\epsilon} \frac{d_{\text{h}}^2 - d_{\text{t}}^2}{r_{ij}^4} e^{-r_{ij}/\lambda}\; \hat{\bm{n}}_j\times \hat{\bm{r}}_{ij},
\end{equation}
where $\ell = 3R/8$ is the distance by which the dipoles are off-centered, with $R=1.5$ $\mu$m the particle radius.

The torque in \cref{eq torque} tends to turn a particle with orientation $\hat{\bm{n}}_j$ away from the interparticle distance vector $\bm{r}_{ij}$ (\cref{Fig interactions}). These turn-away interactions are fundamentally different from Vicsek-type alignment interactions: Whereas alignment interactions couple the orientations of two particles ($\bm{\Gamma}_{ij} \propto \hat{\bm{n}}_i \times \hat{\bm{n}}_j$), our turn-away interactions couple the orientation of one particle to the position of the other one ($\bm{\Gamma}_{ij} \propto \hat{\bm{n}}_j \times \bm{r}_{ij}$). As a result, turn-away torques are intrinsically non-reciprocal: $\bm{\Gamma}_{ij} \neq - \bm{\Gamma}_{ji}$.

We write Langevin equations for the translational and rotational motion of particle $i$ as
\begin{subequations} \label{eq Langevin-main}
\begin{align}
\frac{\dd\bm{r}_i}{\dd t} &= v_0 \hat{\bm{n}}_i(\theta_i) + \frac{\bm{F}_i}{\xi_{\text{t}}} + \sqrt{2D_{\text{t}}}\, \bm{\eta}^{\text{t}}_i(t);\qquad \bm{F}_i = \sum_{j\neq i} \bm{F}_{ji}, \label{eq Langevin-translation-main}\\
\frac{\dd\theta_i}{\dd t} &= \frac{\Gamma_i}{\xi_{\text{r}}} + \sqrt{2D_{\text{r}}}\, \eta_i^{\text{r}}(t);\qquad \Gamma_i= \sum_{j\neq i} \bm{\Gamma}_{ji}\cdot\hat{\bm{z}},\label{eq Langevin-rotation-main}
\end{align}
\end{subequations}
where $v_0$ is the self-propulsion speed, $\hat{\bm{n}}_i = (\cos\theta_i,\sin\theta_i)$ is the orientation of particle $i$, $\xi_{\text{t}}$ and $\xi_{\text{r}}$ are the translational and rotational friction coefficients, respectively, and $\bm{\eta}^{\text{t}}_i(t)$ and $\eta^{\text{r}}_i(t)$ are Gaussian white noises with strengths given by the translational and rotational diffusivities $D_{\text{t}}$ and $D_{\text{r}}$, respectively. We perform Brownian dynamics simulations of this model with $N$ particles in a square box of side $L$ with periodic boundary conditions (\cref{methods}, parameter values in \cref{t parameters}). We benchmark the simulations by reproducing the phase separation reported in Ref. \cite{Zhang2021i}, which is induced by torques that turn particles towards one another, with $d_{\text{t}}^2> d_{\text{h}}^2$ (\cref{Fig turn-towards}).

For turn-away torques, with $d_{\text{h}}^2> d_{\text{t}}^2$, our simulations with $N=2500$ show the emergence of global polar order $P$, as we increase either the self-propulsion speed $v_0$ or the global area fraction $\phi_0 = N \pi R^2/L^2$ (\cref{Fig 2}, \hyperref[movies]{Movies 5 and 6}). Global polar order also emerges in larger simulations of up to $N=70225$ (\cref{Fig system-size}), suggesting that the flocking transition survives in the large-system limit. We conclude that active Brownian particles can flock despite turning away from one another. Moreover, the phase boundary obtained from simulations (\cref{Fig phase-diagram}, blue) is quantitatively close to the transition that we observe in experiments (\cref{Fig experimental-diagram}).

\bigskip

\noindent\textbf{Active Wigner crystals}

\begin{figure*}[tb]
\begin{center}
\includegraphics[width=\textwidth]{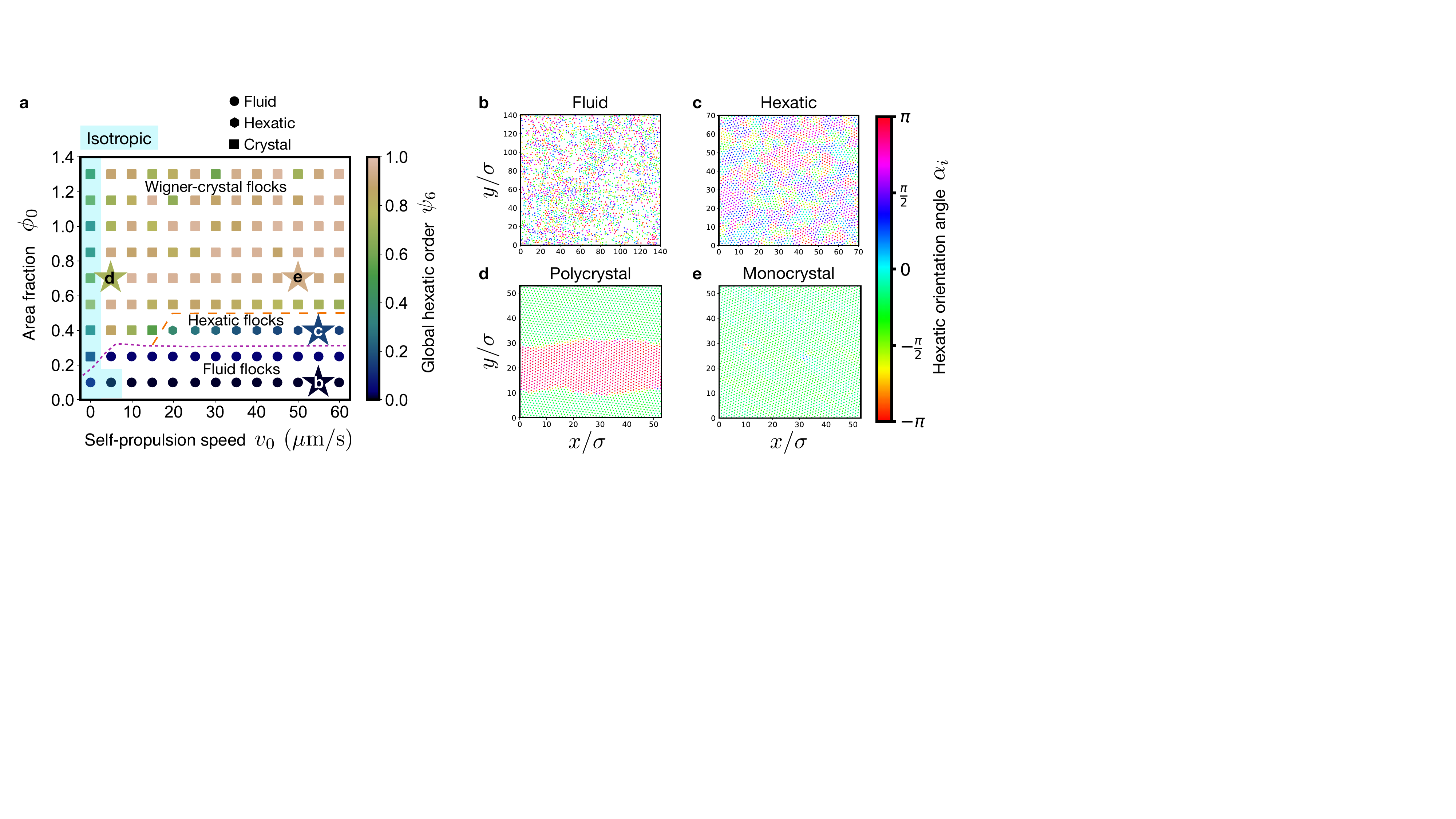}
\end{center}
  {\phantomsubcaption\label{Fig structure}}
  {\phantomsubcaption\label{Fig fluid}}
  {\phantomsubcaption\label{Fig hexatic}}
  {\phantomsubcaption\label{Fig polycrystal}}
  {\phantomsubcaption\label{Fig monocrystal}}
\bfcaption{Active Wigner crystals}{ \subref*{Fig structure}, At high area fraction, the system forms flocking Wigner crystals, which have high values of the global hexatic order parameter $\psi_6$. The hexatic and crystalline phases are identified from orientational and positional correlations (\cref{Fig correlations}), and phase boundaries (dashed lines) are guides to the eye. Stars indicate the snapshots shown in the other panels. \subref*{Fig fluid}-\subref*{Fig monocrystal}, Snapshots of fluid, hexatic, and crystalline flocks. Crystals are typically polycrystalline with grain boundaries at low activity (\subref*{Fig polycrystal}), and monocrystalline at higher activity (\subref*{Fig monocrystal}). Color indicates the angle $\alpha_i$ between the local hexatic order of each particle and its average over the system (\cref{methods}).} \label{Fig 3}
\end{figure*}

As we increase the area fraction in simulations beyond those in our experiments, flocks develop crystalline order (\cref{Fig 3}). These states are reminiscent of flocking crystals reported in previous simulations \cite{Gregoire2003,Menzel2013}. However, whereas ordinary crystals form by attraction between the particles, the force in our model is purely repulsive (\cref{eq force}). Thus, the flocking crystals that we found are active counterparts of Wigner crystals, which were originally proposed to form through electrostatic repulsion in electron gases \cite{Wigner1934}. In our system, the repulsion includes both the electrostatic repulsive force in \cref{eq force} as well as an effective repulsion arising from the turn-away torques in \cref{eq torque} and self-propulsion \cite{Smeets2016,LeBlay2022}. Therefore, turn-away torques promote the formation of active Wigner crystals.

In two dimensions, crystallization involves an intermediate hexatic phase with orientational order in the particle positions \cite{Halperin1978}. We obtain the global hexatic order parameter $\psi_6$, which goes from 0 in the liquid phase to 1 for a monodomain triangular lattice (\cref{Fig structure}, \cref{methods}). We then use orientational and positional correlations to identify the transitions to the hexatic and crystalline phases (\cref{Fig correlations}), marked with dashed lines in \cref{Fig structure}. Consistently with previous works \cite{vanderMeer2016,Ramananarivo2019}, increasing activity $v_0$ in the crystalline phase (\crefrange{Fig polycrystal}{Fig monocrystal}) promotes the formation of a single crystal spanning the entire system. The polycrystalline states at low activity $v_0$ (\cref{Fig polycrystal}) last for the entire duration of our simulations.

Although they both depend on the turn-away torques, the flocking and the crystallization transitions remain separate. Below the crystallization threshold, we observe fluid flocks (\cref{Fig structure,Fig fluid}). Crystallization is therefore not required for flocking via turn-away torques. In fact, as we increase the area fraction in larger systems with $N=40000$ particles, we observe flocks in the form of well-known polar bands and uniform liquids \cite{Bricard2013,Marchetti2013,Chate2020} before reaching hexatic states and active Wigner crystals (\cref{Fig flocking-states}, \hyperref[movies]{Movies 7 to 11}). Overall, crystallization is not involved in the mechanism whereby particles with turn-away interactions flock.

\bigskip

\noindent\textbf{Coarse-graining shows that correlations enable flocking}

\begin{figure*}[tbhp!]
\begin{center}
\includegraphics[width=\textwidth]{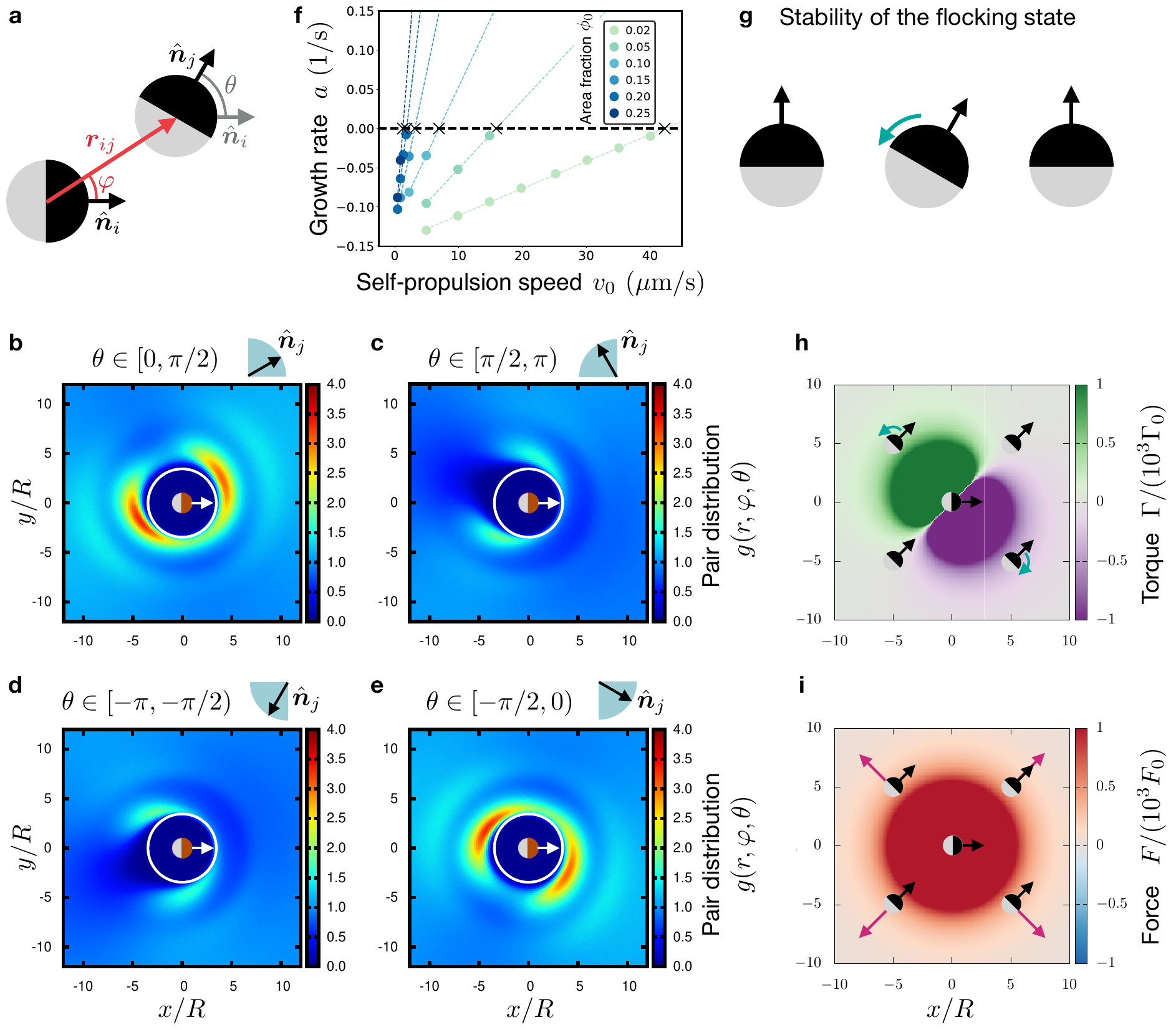}
\end{center}
  {\phantomsubcaption\label{Fig angles}}
  {\phantomsubcaption\label{Fig quadrant-1}}
  {\phantomsubcaption\label{Fig quadrant-2}}
  {\phantomsubcaption\label{Fig quadrant-3}}
  {\phantomsubcaption\label{Fig quadrant-4}}
  {\phantomsubcaption\label{Fig growth-rate}}
  {\phantomsubcaption\label{Fig stability}}
  {\phantomsubcaption\label{Fig torque}}
  {\phantomsubcaption\label{Fig force}}
\bfcaption{Correlations established by turn-away torques and repulsion enable the emergence of polar order}{ \subref*{Fig angles}, Definitions of the arguments $r$, $\varphi$, and $\theta$ of the pair distribution function. \subref*{Fig quadrant-1}-\subref*{Fig quadrant-4}, Pair distribution function measured in simulations in the isotropic state, for $\phi_0 = 0.1$ and $v_0 = 5$ $\mu$m/s, built as histograms over $1000$ independent configurations. The four panels show $g(r,\varphi)$ for particle pairs with relative orientation $\theta$ in each of the four quadrants, as indicated. The white circumferences around the reference particle indicate the predicted exclusion region, which we obtain as the distance $r^*$ at which repulsion overcomes the self-propulsion force: $F(r^*) = \xi_{\text{t}} v_0$, with \cref{eq force} expressed as $\bm{F}_{ij} = F(r) \hat{\bm{r}}_{ij}$. \subref*{Fig growth-rate}, Growth rate calculated using \cref{eq growth-rate} with \cref{eq tau0} using the $g(r,\varphi,\theta)$ measured in simulations. The black crosses, obtained by extrapolation, indicate the onset of flocking, which determine the green phase boundary in \cref{Fig phase-diagram}. \subref*{Fig stability}, Schematic showing that torques stabilize the flocking state: Turning away from both the left and right neighbors keeps particles moving together. \subref*{Fig torque}-\subref*{Fig force}, Torque and force fields, respectively, exerted by the reference particle on a particle with relative orientation $\theta = \pi/4$, as indicated. This orientation belongs to the first quadrant, and hence to panel \subref*{Fig quadrant-1}. Green arrows in \subref*{Fig torque} and purple arrows in \subref*{Fig force} respectively indicate the directions of the torque and repulsive force on each of the probe particles, which explain the pair distribution in panel \subref*{Fig quadrant-1}. The torque and force are plotted from \cref{eq torque,eq force} and normalized by $\Gamma_0 = 3\ell (d_{\text{h}}^2 - d_{\text{t}}^2)/(4\pi\epsilon R^4)$ and $F_0 = 3(d_{\text{h}} + d_{\text{t}})^2/(4\pi\epsilon R^4)$.} \label{Fig 4}
\end{figure*}

To understand the emergence of polar order, we coarse-grain the microscopic model (\cref{eq Langevin-main}). To this end, we write the Smoluchowski equation and break it into the Bogoliubov-Born-Green-Kirkwood-Yvon (BBGKY) hierarchy to obtain an equation for the one-particle distribution function, from which we obtain hydrodynamic equations for the density and the polarity fields $\rho(\bm{r},t)$ and $\bm{P}(\bm{r},t)$ (\cref{derivation}). For the polarity, we obtain
\begin{equation} \label{eq polarity}
\partial_t \bm{P} = a[\rho] \bm{P} + \mathcal{O}\left(\bm{\nabla}\right).
\end{equation}
Polar order emerges if $a > 0$. The coarse-graining yields
\begin{equation} \label{eq growth-rate}
a[\rho] = \frac{\rho}{2\pi \xi_{\text{r}}} \tau_0 - D_{\text{r}},
\end{equation}
where the first term is due to torques, and the second term represents the decay of polar order due to rotational diffusion. The effect of the torques, expressing \cref{eq torque} as $\bm{\Gamma}_{ij} = \Gamma(r) \, \hat{\bm{n}}_j\times \hat{\bm{r}}_{ij}$ with $r= |\bm{r}_{ij}|$, is embodied in the coefficient $\tau_0$, which is given by (\cref{derivation})
\begin{equation} \label{eq tau0}
\tau_0 = \int_0^\infty r\,\dd r \int_0^{2\pi} \dd \varphi \int_0^{2\pi} \dd\theta \, \sin \theta\, \Gamma(r) \sin \varphi\,  g(r,\varphi,\theta)
\end{equation}
in terms of the pair distribution function $g(r,\varphi,\theta)$ in the isotropic state \cite{Grossmann2020}. This function encodes correlations, as it gives the probability density of finding a pair of particles at a distance $r$, with the second particle at a position and orientation forming angles $\varphi$ and $\theta$ with respect to the orientation of the reference particle: $\hat{\bm{n}}_i \cdot \hat{\bm{r}}_{ij} = \cos \varphi$ and $\hat{\bm{n}}_i \cdot \hat{\bm{n}}_j = \cos\theta$ (\cref{Fig angles}).

Which properties must the pair distribution $g$ have in order to yield a non-zero $\tau_0$ that could produce flocking? To prevent the angular integrals in \cref{eq tau0} from vanishing by symmetry \cite{Sese-Sansa2022c}, $g$ has to fulfill the following conditions:
\begin{enumerate}
\item[(i)] $g(r,-\varphi,\theta) \neq g(r,\varphi,\theta)$,
\item[(ii)] $g(r,\varphi,-\theta) \neq g(r,\varphi,\theta)$,
\item[(iii)] $g(r,\varphi + \pi,\theta) \neq g(r,\varphi,\theta)$,
\item[(iv)] $g(r,\varphi,\theta + \pi) \neq g(r,\varphi,\theta)$.
\end{enumerate}
These conditions are necessary, but not sufficient, for flocking.

To test whether these conditions are satisfied, we measure $g(r,\varphi,\theta)$ in our simulations (\cref{methods}). As $g$ is a function of three arguments, we plot $g(r,\varphi)$ and bin the relative orientations in the four quadrants of the angle $\theta$ (\crefrange{Fig quadrant-1}{Fig quadrant-4}). These plots show a clear asymmetry upon changing the sign of the polar angle $\varphi \rightarrow -\varphi$. There is also a clear asymmetry upon the transformation $\theta \rightarrow -\theta$, which corresponds to exchanging the first quadrant with the fourth (\cref{Fig quadrant-1,Fig quadrant-4}) and the second with the third (\cref{Fig quadrant-2,Fig quadrant-3}). Therefore, our system satisfies conditions (i) and (ii). Furthermore, for a given quadrant of $\theta$, changing $\varphi \rightarrow \varphi + \pi$ corresponds to moving to the diametrically-opposed point, which yields a different value of $g$. Respectively, changing $\theta \rightarrow \theta + \pi$ corresponds to exchanging the first quadrant with the third (\cref{Fig quadrant-1,Fig quadrant-3}) and the second with the fourth (\cref{Fig quadrant-2,Fig quadrant-4}), which again yields different values of $g$. Therefore, our system also satisfies conditions (iii) and (iv).

These results show that the correlations in our system fulfil the necessary requirements to yield flocking. We then use the measured pair distribution to predict the flocking transition. To this end, we introduce the measured $g(r,\varphi,\theta)$ into \cref{eq tau0} to obtain the growth rate $a$ of polar order in \cref{eq growth-rate} for different values of the area fraction $\phi_0$ and self-propulsion speed $v_0$ (\cref{Fig growth-rate}). The change of sign of $a$ marks the predicted onset of flocking, shown as the green line in \cref{Fig experimental-diagram,Fig phase-diagram}, which agrees quite well with the experimental results (\cref{Fig experimental-diagram}). Our theory therefore captures the flocking transition.

Once polar order has emerged, turn-away torques stabilize it. In contrast to alignment interactions, turn-away interactions hinder the formation of polar clusters, as particles at the cluster edge turn away and move into low-density areas. Therefore, turn-away interactions produce flocking states in which particles always have lateral neighbors. As a result, if a particle deviates from the flocking direction, it moves closer to a neighbor and experiences a torque that restores its initial orientation (\cref{Fig stability}). Flocking therefore represents a compromise between turning away from left and right neighbors.

\bigskip

\noindent\textbf{Flocking emerges from turn-away torques and repulsion}

Our findings so far indicate that, already in the isotropic state, the system builds up correlations that enable the emergence of polar order. Where do these correlations come from? We argue that they arise from the combined effects of turn-away torques and repulsion forces. Taking \cref{Fig quadrant-1} as an example, the peaks of $g$ are found in regions of low turn-away torque (as shown in \cref{Fig torque}), where particles tend to stay longer. In other quadrants, these low-torque regions change precisely in the way required to satisfy conditions (i) and (ii) (\cref{Fig torques}). Respectively, the asymmetry between the front and the back peaks in \cref{Fig quadrant-1}, required for condition (iii), is due to repulsion. Whereas particles at the front are pushed forwards and sped up by repulsion, particles at the back are pushed back and slowed down (\cref{Fig force}), which makes them stay longer. Finally, condition (iv) is satisfied due to the difference in relative velocity between the particles: The correlations when two particles move in the same direction are different than when they move in opposite directions. We conclude that, together, turn-away torques and repulsion provide the conditions for flocking.

\begin{figure*}[tbh!]
\begin{center}
\includegraphics[width=0.9\textwidth]{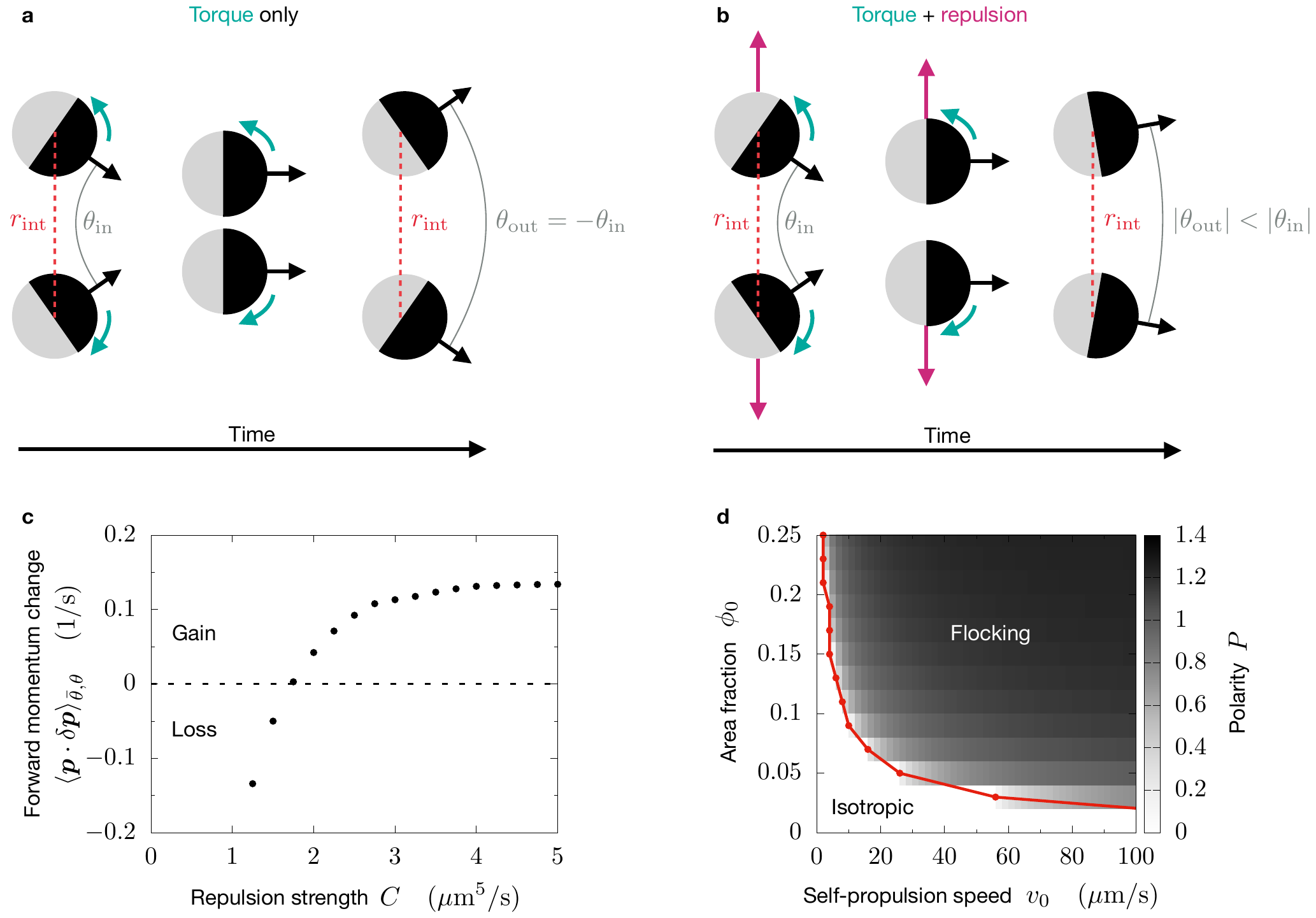}
\end{center}
  {\phantomsubcaption\label{Fig scattering-torque}}
  {\phantomsubcaption\label{Fig scattering-repulsion}}
  {\phantomsubcaption\label{Fig momentum}}
  {\phantomsubcaption\label{Fig kinetic-phase-diagram}}
\bfcaption{Repulsion and turn-away torques produce effective alignment during scattering events}{ \subref*{Fig scattering-torque}-\subref*{Fig scattering-repulsion}, Schematics of a symmetric scattering event, for which turn-away torques alone do not yield any alignment (\subref*{Fig scattering-torque}) whereas adding repulsion does (\subref*{Fig scattering-repulsion}). \subref*{Fig momentum}, Scattering events produce a gain in forward momentum (\cref{eq momentum-change}) only in the presence of enough repulsion. The repulsion strength $C$ is defined from \cref{eq force} by $F(r)= C e^{-r/\lambda}/r^4$ (see \cref{eq scattering} in \cref{methods}). \subref*{Fig kinetic-phase-diagram}, Phase diagram of flocking predicted from the Boltzmann kinetic theory (\cref{eq momentum-change} and \cref{Boltzmann}), The phase boundary is marked in red. The polarity $P$ is obtained from \cref{eq steady-state-polarity} (see \cref{methods}).} \label{Fig 5}
\end{figure*}

\bigskip

\noindent\textbf{Boltzmann kinetic theory predicts effective alignment}

To gain microscopic insight into how torques and repulsion jointly produce flocking, we analyze binary scattering events. We choose the axes so that the two particles are separated along the $\hat{\bm{y}}$ axis (\cref{Fig scattering-geometry}). Then, a scattering event between particles with orientations $\theta_1$ and $\theta_2$ is characterized by the incoming half-angle $\bar\theta = \arg(e^{i\theta_1} + e^{i\theta_2})$ and the angle difference $\theta = \theta_2 - \theta_1$ (\cref{Fig scattering-angles}). We first consider an event with $\bar\theta = 0$, and we analyze scattering with torques only (\cref{Fig scattering-torque}). If particles come into the interaction range $r_{\text{int}}$ at an angle difference $\theta_{\text{in}}$, they will turn away from each other and they will exit the interaction range with angle difference $\theta_{\text{out}} = -\theta_{\text{in}}$ (\cref{Fig scattering-torque}). The momentum of the pair does not change. Therefore, turn-away torques alone do not yield any alignment during this scattering event.

Repulsion, however, pushes particles out of the interaction range before they have time to turn completely (\cref{Fig scattering-repulsion}). Therefore, particles leave the scattering event with a smaller angle difference than initially: $|\theta_{\text{out}}| < |\theta_{\text{in}}|$. The combined effects of turn-away torques and repulsion therefore cause effective alignment, and hence they can produce flocking.

To average over scattering events, we use Boltzmann's kinetic theory generalized for self-propelled particles \cite{Lam2015}. This theory predicts that the growth rate of polar order, as in \cref{eq polarity}, is given by $a = \left\langle \bm{p}\cdot \delta\bm{p}\right\rangle_{\bar\theta,\theta} - D_{\text{r}}$, where the first term is the average over scattering events of the dimensionless momentum change $\delta \bm{p}$ in the forward direction, i.e., projected on the incoming momentum $\bm{p} = \hat{\bm{n}}_1 + \hat{\bm{n}}_2$. In our case, it is given by (\cref{Boltzmann})
\begin{equation} \label{eq momentum-change}
\left\langle \bm{p}\cdot \delta \bm{p}\right\rangle_{\bar\theta,\theta} = \frac{\rho v_0 r_{\text{int}}}{2\pi} \int_{-\pi}^\pi \dd \theta \left| \sin\left(\theta/2\right)\right| \left\langle \bm{p}\cdot \delta\bm{p}(\bar\theta,\theta)\right\rangle_{\bar\theta},
\end{equation}
where $r_{\text{int}}=\lambda$ is the interaction range. This forward momentum change quantifies the effective alignment arising from scattering events. To obtain it, we numerically integrate the equations of motion \cref{eq Langevin-main} without noise for binary scattering events with different $\bar\theta$ (\cref{methods}). The results show that having forward momentum gain requires repulsion (\cref{Fig momentum}), which is therefore necessary for flocking. Finally, we use the scattering statistics to predict the growth rate $a$, and hence the onset of flocking for $a=0$ (\cref{Fig kinetic-phase-diagram}, red). These predictions, which include only two-particle dynamics, are consistent with our many-particle simulation results and with our experiments (\cref{Fig experimental-diagram}). Together, these results reveal the microscopic mechanism responsible for flocking through combined repulsion and turn-away torques.

\bigskip

\noindent\textbf{Discussion and outlook}

In summary, we discovered a mechanism that allows self-propelled particles to flock by repelling and turning away from each other. This finding sharply contrasts with the mechanism of flocking in the paradigmatic Vicsek model, which relies on explicit alignment interactions between the active agents. In that case, flocking emerges purely from torques; it does not require any central forces between the agents. In contrast, we revealed a mechanism of flocking that relies on both torques and repulsion, thus combining features of aligning and non-aligning active matter.

Recent work reported that flocking can arise from collision-avoidance interactions, which align particles by turning them away from a collision \cite{Grossmann2020,Chen2023a}. Our turn-away interactions, however, can produce severe misalignment, as particles keep turning away from each other even after they have avoided collision, thus making flocking seemingly impossible. Our results show that flocking emerges even in this case.

At high densities and speeds, alignment-based flocks made of Quincke rollers were previously found to lose polar order as they experience motility-induced phase separation (MIPS) \cite{Geyer2019}. However, the turn-away torques in our system hinder MIPS, as they reorient particles away from clusters \cite{Knezevic2022}. Therefore, flocks obtained through turn-away interactions can achieve higher speeds and densities without suffering phase separation and loss of collective motion. This feature allows us to obtain flocks in the form of dense liquids and Wigner crystals produced by particle repulsion. Similar crystals, albeit without flocking, were obtained recently in simulations of microswimmers with chemorepulsive interactions \cite{Yang2024}.

Active particles often crystallize via either MIPS \cite{Redner2013,Palacci2013,VanderLinden2019,Omar2021}, hydrodynamic attraction \cite{Singh2016,Thutupalli2018,Bililign2022,Tan2022a}, or by approaching close packing \cite{Bialke2012,Weber2014,Briand2016}, all of which produce particle collisions and impair collective motion. Here, Wigner crystallization keeps particles at a distance, which avoids collisions and does not impair collective motion.

As an outlook, we speculate that flocking by turning away might contribute to collective motion in cell populations, possibly solving the apparent paradox that cells can flock despite interacting via contact inhibition of locomotion \cite{Smeets2016,Hiraiwa2020,Bertrand2024}. Interactions between cells are much more complex than those between our active colloids. Yet, as mesenchymal cells repolarize away from each other upon collision, cell-cell scattering events could have outcomes similar to those between our particles, with an outgoing angle larger than zero but smaller in magnitude than the incoming angle (\cref{Fig scattering-repulsion}). We look forward to experimental tests of this idea in future work.

Finally, flocking by turning away might also provide a strategy to engineer robust swarming in disordered environments. We already saw hints of this robustness in our experiments, in which flocks easily flow past stuck particles (\hyperref[movies]{Movies 2 and 4}). If an agent gets stuck at an obstacle, alignment interactions will produce accumulation of followers behind it. Instead, turn-away interactions allow followers to readily reorient away, which might yield smoother and more efficient flocking through disordered landscapes.

\pagebreak


\noindent\textbf{Acknowledgments}

We thank Jonathan Bauermann, Fridtjof Brauns, Kartik Chhajed, Erwin Frey, Steve Granick, Robert Gro\ss man, Frank J\"{u}licher, Debasmit Sarkar, Jeanine Shea, Holger Stark, John Toner, and Till Welker for discussions. We thank Mariona Esquerda Ciutat for Fig. 1a. We acknowledge computing support by the Max Planck Computing and Data Facility and the computing facility at MPI-PKS. Z. Z. and J. Z. acknowledge financial support from the National Natural Science Foundation of China (NSFC) (Grant No. 12204453) and University of Science and Technology of China (YD2060002028, WK3450000008, KY2060000211). Z. Z. and J. Z. acknowledge the Supercomputing Center of University of Science and Technology of China (USTC) for  computational resources. Experimental work was partially carried out at the Center for Micro and Nanoscale Research and Fabrication and the Instruments Center for Physical Science, University of Science and Technology of China.


\bigskip

\noindent\textbf{Author contributions}

S.D. performed and analyzed the Brownian dynamics simulations. S.D. and R.A. developed the BBGKY hierarchy calculations. M.C. performed the scattering calculations and simulations. J.Y. performed the initial set of experiments during his PhD. J.Z. performed additional experiments. S.D., Z.Z., J.Y., and J.Z. analyzed the experimental data. R.A. conceived and supervised the work. S.D., M.C., and R.A. wrote the paper.




\bigskip

\noindent\textbf{Data and code availability}

Data and code are available \MYhref{https://github.com/dassuchi/Flocking}{on this link}.

\bibliography{Flocking}

\onecolumngrid 

\clearpage
\appendix

\twocolumngrid

\section{Methods} \label{methods}

\subsection{Particle synthesis} \label{synthesis}

Following protocols described elsewhere \cite{Yan2016}, a submonolayer of $3$ $\mu$m-diameter silica particles (Tokuyama) is prepared on a standard glass slide. Then, $35$ nm of titanium and then $20$ nm of SiO$_2$ are deposited vertically on the glass slide using an electron-beam evaporator. The preparation is then washed with isopropyl alcohol and deionized water, and then sonicated into deionized water to collect the Janus particles.

\subsection{Experimental setup} \label{setup}

The particle suspensions are confined between two coverslips (SPI Supplies) coated with indium tin oxide to make them conductive, and with $25$ nm of silicon oxide to prevent particles from sticking to them. The coverslips have a $9$ mm hole in the center, separated by a $120$ $\mu$m-thick spacer (GraceBio SecureSeal), where we place the suspension of Janus colloidal particles. An alternating voltage is applied between the coverslips using a function generator (Agilent 33522A). The sample cell is imaged with 5$\times$, 40$\times$, and 64$\times$ air objectives on an inverted microscope (Axiovert 200). The observation areas are $1232$ $\mu$m $\times$ $1640$ $\mu$m, $154$ $\mu$m $\times$ $205$ $\mu$m, and $96$ $\mu$m $\times$ $128$ $\mu$m respectively. Microscopic images and videos are taken with a CMOS camera (Edmund Optics 5012M GigE) with $20$ ms time resolution.

\subsection{Image analysis} \label{image-analysis}

Image processing is performed using MATLAB with home-developed codes.

\subsection{Experimental data analysis} \label{data-analysis}

Here we describe the analysis methods that we use to identify flocking in the experimental movies. We analyze experimental movies obtained with three different magnifications: 64$\times$, 40$\times$, and 5$\times$.

For 64$\times$ and 40$\times$ magnifications, we can track particle orientations $\hat{\bm{n}}_i(t)$ and obtain the polar order parameter $P(t) = \frac{1}{N} | \sum_{i=1}^N {\hat{\bm{n}}_i}(t) |$ (\cref{Fig experiment-polar-order}). We classify as flocking the experimental realizations for which the time-averaged polar order parameter is $\left\langle P(t)\right\rangle_t \geq 0.5$.

For 5$\times$ magnification, we cannot resolve single-particle orientations $\hat{\bm{n}}_i(t)$, and hence we can only study polar order indirectly through the flow field $\bm{v}(\bm{r},t)$, which we measure using particle image velocimetry (PIV). Analyzing high-magnification videos, we found that the polar order parameter obtained from particle orientations, $P_n (t) = P(t) = \frac{1}{N} | \sum_{i=1}^N {\hat{\bm{n}}_i}(t) |$ strongly correlates with that obtained from the particle velocity axes $\hat{\bm{v}}_i(t)$ as $P_v(t) = \frac{1}{N} | \sum_{i=1}^N {\hat{\bm{v}}_i}(t) |$ (\cref{Fig velocity-polarity-correlation}). Hence, the velocity field approximates well the polarity field, and we therefore use it to characterize polar order in low-magnification movies.

To do this, we obtain the spatial correlation function of the velocity field, $C(r,t) = \left\langle \bm{v}(r,t) \cdot \bm{v}(0,t)\right\rangle$. We then fit to it an exponential decay $C(r,t) \sim e^{-r/\xi(t)}$, from which we extract the velocity correlation length $\xi(t)$ (\cref{Fig correlation-length}). We expect flocking states to have a correlation length substantially larger than the average interparticle distance $\left\langle r\right\rangle = \sqrt{1/\rho} = R\sqrt{\pi/\phi_0}$. We classify as flocking the experimental realizations for which the time-averaged correlation length is larger than 10 times the average interparticle distance: $\left\langle \xi(t)\right\rangle_t \geq 10\left\langle r \right\rangle_{\bm{r},t}$ (\cref{Fig correlation-length-analysis}). This threshold seemed to be a good compromise between requiring relatively long-ranged correlations and accounting for the finite size of the polar domains throughout our experiments. In most movies that visually exhibit some polar order, the polar domains are not much bigger than this threshold size. Therefore, the threshold cannot be increased much further. At the same time, setting a lower threshold would classify as flocking some experimental realizations in which polar order is visually inexistent or only very short-ranged.

\subsection{Simulation scheme} \label{numerical}

We implement Brownian dynamics simulations of \cref{eq Langevin-main} with \cref{eq force,eq torque}. We place the particles in a square box of side $L$ with periodic boundary conditions. We use simulation units and parameter values estimated from our experiments as indicated in \cref{t parameters}. We use an explicit Euler-Mayurama method for the time evolution with time step $\Delta t = 10^{-4}$, and the simulations are performed for $N_{\text{steps}} = 1.4\times 10^6$ steps, corresponding to a simulation duration $T \approx 934$ s.

We benchmark our simulations by reproducing the results of Ref. \cite{Zhang2021i} using $N=1024$ particles with turn-towards torques (\cref{Fig turn-towards}). For turn-away torques, we perform simulations with $N=2500$ particles. We show non-uniform flocking states in larger systems using simulations with $N=40000$ particles (\cref{Fig flocking-states}). In this case, we use a cell-list algorithm to track particle neighbors. In our simulations, we vary the area fraction $\phi_0 = N\pi \sigma^2/(4L^2)$ and the self-propulsion speed $v_0$ in the ranges $\phi_0 \in \left[0.02, 1.30\right]$ and $v_0\in \left[0, 60\right] \mu$m/s.

\begin{table*}[!htb]
\begin{center}
\begin{tabular}{lcc}

Description and symbol&Estimate&Value in simulation units\\\hline

Particle diameter $\sigma = 2R$&$3$ $\mu$m&$1$\\
Rotational diffusion coefficient $D_{\text{r}}$&$0.15$ s$^{-1}$&$1$\\
Thermal energy $k_{\text{B}} T$&$4.11\times 10^{-21}$ J&$1$\\
Dielectric permittivity of the solvent $\epsilon$&$6.95\times 10^{-10}$ C$^2$/(N m$^2$)&$1$\\
Translational diffusion coefficient $D_{\text{t}}$&$0.16$ $\mu$m$^2$/s&$0.11$\\
Translational drag coefficient $\xi_{\text{t}}$&$25$ mPa$\cdot$s$\cdot\mu$m&$9$\\
Rotational drag coefficient $\xi_{\text{r}}$&$75$ mPa$\cdot$s$\cdot\mu$m$^3$&$3$\\
Electrostatic screening length $\lambda$&$120$ $\mu$m&$40$\\
Dipole shift distance $\ell$&$0.56$ $\mu$m&$0.19$\\
Electric field amplitude $E_0$&$83$ V/mm&$179$\\
Head dipole magnitude $|d_{\text{h}}|$&$9.76\times 10^{-22}$ C$\cdot$m&$112$\\
Tail dipole magnitude $|d_{\text{t}}|$&$4.64\times 10^{-22}$ C$\cdot$m&$53$\\
\end{tabular}
\bfcaption{Parameter estimates}{ Parameter values as obtained in Ref. \cite{Zhang2021i}, except switching the head and tail dipole magnitudes. We define our simulation units based on the four first entries of this table, which we use to define the scales of length, time, energy, and electric charge.} \label{t parameters}
\end{center}
\end{table*}

\subsection{Quantification of hexatic order} \label{hexatic}

To quantify hexatic order \cite{Halperin1978}, we measure the local hexatic order of particle $i$ defined as $\psi_{6,i} = \frac{1}{N_{\text{nn}}^i} \sum_{j= 1}^{N_{\text{nn}}^i} e^{6i\theta_{ij}}$, where $\theta_{ij}$ is the angle of the segment connecting particles $i$ and $j$ with respect to the $\hat{\bm{x}}$ axis, and $N_{\text{nn}}^i$ is the number of nearest neighbours of particle $i$ found through Voronoi tessellation. From it, we obtain the global hexatic order $\psi_6 = \frac{1}{N} |\sum_{i = 1}^{N} \psi_{6,i}|$, which is a scalar order parameter that we show in \cref{Fig structure}. We also obtain the hexatic angle $\alpha_i$ of each particle, which we show to visualize ordered domains in \crefrange{Fig fluid}{Fig monocrystal}. This angle indicates the projection of the local hexatic order $\psi_{6,i}$ onto its average $\Psi_6 = \frac{1}{N} \sum_{i=1}^N \psi_{6,i}$, obtained from the scalar product of these two complex numbers: $\psi_{6,i} \Psi_6 = \left|\psi_{6,i}\right| \left|\Psi_6\right| \cos \alpha_i$.

\subsection{Pair distribution function} \label{gr}


To numerically obtain the pair distribution function $g(r,\varphi,\theta)$, we build a histogram of the local number of particles $N_l(r, \varphi, \theta)$ at distance $r$, positional angle $\varphi$, and relative orientation $\theta$ (\cref{Fig angles}). We normalise the count to obtain $g(r, \varphi, \theta) = 2\pi N_l(r, \varphi, \theta)/(A(r) N t_{\text{run}} \rho \Delta \theta)$, where $A(r)= r\,\Delta r\,\Delta\varphi$ is the area of the annular segment of radial width $\Delta r$ and angular width $\Delta \varphi$, $\rho$ is the number density, $t_{\text{run}}$ is the number of snapshots, and $\Delta \theta$ is the size of the relative-orientation bins \cite{Allen2017}. The histogram is built over $1000$ independent realisations, with a total of $93289$ snapshots. We chose bins of $\pi/180$ for $\varphi$, $0.1$ for $r$, and $\pi/2$ for $\theta$, as shown in \crefrange{Fig quadrant-1}{Fig quadrant-4}.

\subsection{Scattering simulations} \label{scattering}

We initialized binary scattering events by placing the two particles at a distance $|\bm{r}_{12}| = r_{\text{int}}$, when they enter the interaction range (\cref{Fig scattering-torque,Fig scattering-repulsion}). We chose the axes so that the two particles are initially separated along the $\hat{\bm{y}}$ axis (\cref{Fig scattering-geometry}). The direction of the interparticle distance vector will in general vary throughout the scattering event (\cref{Fig scattering-geometry}). For convenience, in these simulations we describe the particle orientations via the angles $\theta_1$ and $\theta_2$ measured with respect to the axis perpendicular to the interparticle distance vector (\cref{Fig scattering-geometry}). Based on these angles, the initial configuration is characterized by the incoming half-angle $\bar\theta = \arg(e^{i\theta_1} + e^{i\theta_2})$ and the angle difference $\theta = \theta_2 - \theta_1$ (\cref{Fig scattering-angles}).

To simulate binary scattering events, we rewrite \cref{eq Langevin-main} without noise in terms of the angles $\theta_1$ and $\theta_2$ and the interparticle distance $r=|\bm{r}_{12}|$ as
\begin{subequations} \label{eq scattering}
\begin{align}
\dot{r} & = \frac{C}{r^4}e^{-r/\lambda} - v_0 \left(\sin{\theta_1} - \sin{\theta_2} \right), \label{eq r} \\
\dot{\theta}_1 &= -\frac{1}{r^4}e^{-r/\lambda} \cos{\theta_1} - \frac{v_0}{r} \left(\cos{\theta_1} - \cos{\theta_2} \right), \label{eq theta1} \\
\dot{\theta}_2 &= \frac{1}{r^4}e^{-r/\lambda} \cos{\theta_2} - \frac{v_0}{r} \left(\cos{\theta_1} - \cos{\theta_2} \right), \label{eq theta2}
\end{align}
\end{subequations}
where $C$ is a parameter that captures the repulsion strength in \cref{eq force} relative to that of torques in \cref{eq torque}. The second term in \cref{eq theta1,eq theta2} is the geometric contribution that accounts for the variation of the interparticle distance vector $\bm{r}_{12}$ as particles move.

We then sample the initial angles $\theta_1(t=0)$ and $\theta_2(t=0)$ from the interval $[-\pi,\pi]$ in steps of $2\pi/100$. For each scattering configuration, we numerically evolve the distance and orientations of the particles according to \cref{eq scattering} until the distance $r$ reaches $r_{\text{int}}$ again and the particles point away from each other, so that they leave the interaction range. For each of these evolutions, we measure the momentum change $\delta\bm{p} = \hat{\bm{n}}'_1 + \hat{\bm{n}}'_2 - \hat{\bm{n}}_1 - \hat{\bm{n}}_2$, where primes indicate the final state, and we use it to numerically perform the integral in \cref{eq momentum-change} to obtain the average forward momentum change $\left\langle \bm{p}\cdot \delta \bm{p}\right\rangle_{\bar\theta,\theta}$.

From this quantity, we calculate the growth rate of the polarity, given by $a = \left\langle \bm{p}\cdot \delta \bm{p}\right\rangle_{\bar\theta,\theta} - D_{\text{r}}$. When $a<0$, the isotropic state is linearly stable, and the polarity vanishes, $P=0$. When $a>0$, the isotropic state is linearly unstable and it gives rise to flocking. To calculate the steady-state polarity of the flocking state, shown in \cref{Fig kinetic-phase-diagram}, we expand the polarity equation to third order as
\begin{equation} \label{eq polarity-third-order}
\partial_t P = a P - b P^3,
\end{equation}
and we calculate the prefactor of the third-order term, which is given by \cite{Lam2015}
\begin{equation} \label{eq third-order}
b = \left\langle \left(1/2 - \cos\theta\right) \bm{p}\cdot \delta \bm{p}\right\rangle_{\bar\theta,\theta}.
\end{equation}
We obtained $b>0$. Therefore, in the range of parameters that we explored, the system undergoes a second-order transition \cite{Lam2015}. Finally, we obtain the steady-state polarity as
\begin{equation} \label{eq steady-state-polarity}
P = \sqrt{a/b}.
\end{equation}

To show that repulsion is required for flocking via turn-away interactions, we varied the repulsion strength $C$ in \cref{eq r} from $0$ up to the value obtained from the experimental estimates in \cref{t parameters}. We then calculated $\left\langle \bm{p}\cdot \delta \bm{p}\right\rangle_{\bar\theta,\theta}$ for each value of $C$ (\cref{Fig momentum}).









\section{Coarse-graining: From the microscopic model to a hydrodynamic description} \label{derivation}


Here, we provide a systematic coarse-graining of the microscopic equations of motion (\cref{eq Langevin-main}) to derive the growth rate of the polarity field $\bm{P}$ (\crefrange{eq polarity}{eq tau0}), which determines the onset of flocking. To this end, we combine and adapt the derivations in Refs. \cite{Zhang2021i,Grossmann2020}.

We start with the Smoluchowski equation, which governs the evolution of the full $N$-particle distribution function of the system $\Psi_N (\bm{r}_1, \hat{\bm{n}}_1, \ldots,\bm{r}_N, \hat{\bm{n}}_N; t)$. We then break it into the BBGKY hierarchy of equations for the 1,2,3,...-particle distribution functions. We truncate the hierarchy at the 2-particle order, with pair correlations encoded in integrals known as the collective force and torque. We then perform a gradient expansion of these integrals, which thereby become local terms in the hydrodynamic equations. Finally, we obtain the hydrodynamic equations by defining continuum fields like the density and the polarity fields as moments of the one-particle distribution function $\Psi_1 (\bm{r}_1,\hat{\bm{n}}_1;t)$.

\subsection{Smoluchowski equation and the BBGKY hierarchy}

The behavior of the system encoded in the set of coupled Langevin equations \cref{eq Langevin-main} can be equivalently described by the Smoluchowski equation for the N-particle distribution function $\Psi_N(\bm{r}_1,\hat{\bm{n}}_1,\ldots,\bm{r}_N,\hat{\bm{n}}_N;t)$, which is the probability density of finding the $N$ particles at positions $\bm{r}_1,\ldots,\bm{r}_N$ with orientations $\hat{\bm{n}}_1,\ldots,\hat{\bm{n}}_N$ at time $t$:
\begin{equation} \label{eq N-particle}
\partial_t \Psi_N = - \sum_{i=1}^N \left[ \bm{\nabla}_i \cdot \bm{J}_{\text{t},i} + \hat{\bm{n}}_i\times \partial_{\hat{\bm{n}}_i} \cdot \bm{J}_{\text{r},i}\right].
\end{equation}
Here, $\hat{\bm{n}}\times \partial_{\hat{\bm{n}}}$ is the rotation operator, and $\bm{J}_{\text{t}}$ and $\bm{J}_{\text{r}}$ are the translational and rotational probability currents, respectively, given by
\begin{subequations}
\begin{align}
\bm{J}_{\text{t},i} &= \left[ v_0\hat{\bm{n}}_i + \frac{\bm{F}_i}{\xi_{\text{t}}}\right] \Psi_N - D_{\text{t}}\,\bm{\nabla}_i\Psi_N,\\
\bm{J}_{\text{r},i} &= \frac{\bm{\Gamma}_i}{\xi_{\text{r}}}\Psi_N - D_{\text{r}}\,\hat{\bm{n}}_i\times \partial_{\hat{\bm{n}}_i}\Psi_N.
\end{align}
\end{subequations}
Hereafter, we describe the two-dimensional particle orientation $\hat{\bm{n}}_i$ in terms of the angle $\theta_i$: $\hat{\bm{n}}_i = (\cos\theta_i,\sin\theta_i)^T$.

Integrating over the positions and orientations of all particles but one, we obtain an equation for the one-particle distributon function $\Psi_1(\bm{r}_1,\theta_1;t)$:
\begin{multline} \label{eq psi1}
\partial_t \Psi_1 = -\bm{\nabla}_1\cdot\left[\left( v_0\hat{\bm{n}}_1 - D_{\text{t}}\bm{\nabla}_1\right) \Psi_1\right] - \bm{\nabla}_1 \cdot \frac{\bm{F}_{\text{int}}}{\xi_{\text{t}}} \\
 - \partial_{\theta_1} \frac{\hat{\bm{z}}\cdot\bm{\Gamma}_{\text{int}}}{\xi_{\text{r}}} + D_{\text{r}}\, \partial_{\theta_1}^2 \Psi_1.
\end{multline}
Here, $\bm{F}_{\text{int}}$ and $\bm{\Gamma}_{\text{int}}$ are the collective force and torque, respectively, which encode the effects of interactions on particle 1. For pair-wise interactions, the collective force and torque can be expressed in terms of the two-particle distribution function $\Psi_2 (\bm{r}_1,\theta_1,\bm{r}_2,\theta_2;t)$ as
\begin{subequations} \label{eq collective-psi2}
\begin{align}
&\begin{multlined}
\bm{F}_{\text{int}}(\bm{r}_1,\theta_1;t) =\\
 -\int \dd^2\bm{r}' \, \dd\theta' \; F(\left| \bm{r}'-\bm{r}_1\right|)\, \frac{\bm{r}'-\bm{r}_1}{\left| \bm{r}'-\bm{r}_1\right|} \,\Psi_2(\bm{r}_1,\theta_1,\bm{r}',\theta';t),
 \end{multlined}
\\
&\begin{multlined}
\bm{\Gamma}_{\text{int}}(\bm{r}_1,\theta_1;t) =\\
 \int \dd^2\bm{r}'\, \dd\theta' \; \Gamma(\left| \bm{r}'-\bm{r}_1\right|)\, \hat{\bm{n}}_1 \times \frac{\bm{r}'-\bm{r}_1}{\left| \bm{r}'-\bm{r}_1\right|}\, \Psi_2(\bm{r}_1,\theta_1,\bm{r}',\theta';t),
\end{multlined}
\end{align}
\end{subequations}
Here,
\begin{subequations} \label{eq force-torque-fields-expressions}
\begin{align}
F(r) &= \frac{3(d_{\text{h}} + d_{\text{t}})^2}{4\pi\epsilon} \frac{e^{-r/\lambda}}{r^4}, \label{eq force-field}\\
\Gamma(r) &= \frac{3\ell(d_{\text{h}}^2 - d_{\text{t}}^2)}{4\pi\epsilon} \frac{e^{-r/\lambda}}{r^4} \label{eq torque-field}
\end{align}
\end{subequations}
are the scalar magnitudes of the interaction force and torque in \cref{eq force,eq torque}.

With the collective force and torque given by \cref{eq collective-psi2}, \cref{eq psi1} is an integro-differential equation for $\Psi_1$ that involves $\Psi_2$. Therefore, \cref{eq psi1} is the first equation in the BBGKY hierarchy. To truncate the hierarchy, we decompose $\Psi_2$ as
\begin{multline}
\Psi_2(\bm{r}_1,\theta_1,\bm{r}',\theta';t) =\\
\Psi_1(\bm{r}',\theta';t) g(\bm{r}',\theta' | \,\bm{r}_1,\theta_1;t) \Psi_1(\bm{r}_1,\theta_1;t).
\end{multline}
Here, $\Psi_2(\bm{r}_1,\theta_1,\bm{r}',\theta';t)$ is the density of particle pairs with one particle at position $\bm{r}_1$ with orientation $\theta_1$ and another particle at position $\bm{r}'$ with orientation $\theta'$ at time $t$. Respectively, $g$ is the dimensionless pair distribution function that encodes the conditional probability of finding a particle at position $\bm{r}'$ and orientation $\theta'$ given that another particle is at position $\bm{r}_1$ with orientation $\theta_1$. Introducing this decomposition into \cref{eq psi1} allows us to express it as a closed equation for $\Psi_1$, hence closing the hierarchy. This closure goes beyond the molecular chaos approximation, as it keeps information about pair correlations in the pair distribution function $g$.

In homogeneous steady states, the probability distributions are time-independent, and the pair correlations do not depend on the coordinates of a given particle but only on the relative coordinates of particle pairs. Hence, we express $g$ in terms of the distance $|\bm{r}'-\bm{r}_1|$ between particles, the angle $\varphi$ formed between the interparticle distance vector $\bm{r}'-\bm{r}_1$ and the orientation vector $\hat{\bm{n}}_1$ of particle 1, defined by $\hat{\bm{n}}_1\cdot \left(\bm{r}'-\bm{r}_1\right) = |\bm{r}'-\bm{r}_1| \cos\varphi$, and the relative orientation $\theta'-\theta_1$. Therefore, in homogeneous steady states like the isotropic state of the Janus particle system that we analyze, we have
\begin{equation}
g(\bm{r}', \theta' | \,\bm{r}_1,\theta_1;t) = g(|\bm{r}'-\bm{r}_1|,\varphi, \theta'-\theta_1).
\end{equation}
With this decomposition, the collective force and torque (\cref{eq collective-psi2}) are expressed as
\begin{subequations} \label{eq collective-g}
\begin{align}
&\begin{multlined} \label{eq collective-force-g}
\bm{F}_{\text{int}}(\bm{r}_1,\theta_1) =  -\Psi_1(\bm{r}_1,\theta_1) \int \dd^2\bm{r}' \,\dd\theta' \; F(\left| \bm{r}'-\bm{r}_1\right|)\\
\frac{\bm{r}'-\bm{r}_1}{\left| \bm{r}'-\bm{r}_1\right|} \,\Psi_1(\bm{r}',\theta') \, g(|\bm{r}'-\bm{r}_1|,\varphi,\theta'-\theta_1),
\end{multlined}
\\
&\begin{multlined} \label{eq collective-torque-g}
\bm{\Gamma}_{\text{int}}(\bm{r}_1,\theta_1) =  \Psi_1(\bm{r}_1,\theta_1) \int \dd^2\bm{r}' \,\dd\theta'\; \Gamma(\left| \bm{r}'-\bm{r}_1\right|)\\
\hat{\bm{n}}_1 \times \frac{\bm{r}'-\bm{r}_1}{\left| \bm{r}'-\bm{r}_1\right|}\, \Psi_1(\bm{r}',\theta')\, g(|\bm{r}'-\bm{r}_1|,\varphi,\theta'-\theta_1).
\end{multlined}
\end{align}
\end{subequations}

\subsection{Gradient expansion}

The collective force and torque in \cref{eq collective-g} depend non-locally on the one-particle distribution function $\Psi_1(\bm{r}',\theta')$. To derive local hydrodynamic equations, we perform a gradient expansion around $\bm{r}_1$:
\begin{equation} \label{eq gradient-expansion}
\Psi_1(\bm{r}',\theta')\approx \Psi_1(\bm{r}_1,\theta') + \bm{\nabla}_{\bm{r}'}\Psi_1(\bm{r}_1,\theta') \cdot(\bm{r}'-\bm{r}_1).
\end{equation}
We also expand in angular Fourier modes:
\begin{equation} \label{eq Fourier-expansion}
\Psi_1 (\bm{r}',\theta') = \frac{1}{2\pi} \sum_{k=0}^\infty e^{-ik\theta'} \,\tilde{\Psi}_{1,k}(\bm{r}').
\end{equation}
Thus, to lowest order in the gradient expansion, we have
\begin{equation} \label{eq gradient-Fourier-expansion}
\Psi_1 (\bm{r}',\theta') \approx \frac{1}{2\pi} \sum_{k=0}^\infty e^{-ik\theta'} \,\tilde{\Psi}_{1,k}(\bm{r}_1).
\end{equation}

Introducing \cref{eq gradient-Fourier-expansion} into \cref{eq collective-torque-g}, and changing the integration variables to the relative coordinates $\bm{r}\equiv \bm{r}'-\bm{r}_1$, with polar coordinates $(r,\varphi)$, and $\theta\equiv \theta'-\theta_1$, we obtain the lowest-order contribution to the collective torque:
\begin{multline}
\bm{\Gamma}^{(0)}_{\text{int}}(\bm{r}_1,\theta_1) = \Psi_1(\bm{r}_1,\theta_1) \int_0^\infty \dd r \; r\, \Gamma(r) \int_0^{2\pi} \dd\varphi \sin\varphi \,\hat{\bm{z}} \\
\times \int \dd\theta \; g(r,\varphi,\theta) \frac{1}{2\pi} \sum_{k=0}^\infty e^{-ik(\theta + \theta_1)} \, \tilde\Psi_{1,k}(\bm{r}_1)\\
= \Psi_1(\bm{r}_1,\theta_1) \frac{1}{2\pi} \sum_{k=0}^\infty e^{-ik\theta_1} \int_0^\infty \dd r \; r\, \Gamma(r) \int_0^{2\pi} \dd\varphi \sin\varphi \,\hat{\bm{z}}\\
\times \int \dd\theta \; g(r,\varphi,\theta) e^{-ik \theta} \, \tilde\Psi_{1,k}(\bm{r}_1).
\end{multline}
Here, we have used that $\hat{\bm{n}}_1\times \hat{\bm{r}} = \sin \varphi \,\hat{\bm{z}}$. Gathering all the integrals into a coefficient, we rewrite the collective torque as
\begin{equation} \label{eq collective-torque-lowest-coefficients}
\bm{\Gamma}^{(0)}_{\text{int}}(\bm{r}_1,\theta_1) = \Psi_1(\bm{r}_1,\theta_1) \frac{1}{2\pi} \sum_{k=0}^\infty e^{-ik\theta_1} \tilde\Psi_{1,k}(\bm{r}_1) \,\tau_{0,k} \,\hat{\bm{z}},
\end{equation}
where we have defined the coefficient
\begin{equation} \label{eq torque-coefficients-exp}
\tau_{0,k} = \int_0^\infty \dd r \; r\, \Gamma(r) \int_0^{2\pi} \dd\varphi \sin\varphi \int_0^{2\pi} \dd\theta \; g(r,\varphi,\theta) e^{-ik \theta}.
\end{equation}
As shown in \crefrange{Fig quadrant-1}{Fig quadrant-4}, the pair distribution $g$ is invariant upon simultaneous inversion of the angles $\varphi$ and $\theta$: $g(r,\varphi,\theta) = g(r,-\varphi,-\theta)$. This symmetry implies that the real part of \cref{eq torque-coefficients-exp} vanishes, and hence only the imaginary part survives:
\begin{multline} \label{eq torque-coefficients}
\tau_{0,k} =  -i \int_0^\infty \dd r \; r\, \Gamma(r)\\
\times \int_0^{2\pi} \dd\varphi \sin\varphi \int_0^{2\pi} \dd\theta\; g(r,\varphi,\theta) \sin(k \theta).
\end{multline}

To complete the gradient expansion, we express \cref{eq psi1} only to lowest order in gradients:
\begin{equation} \label{eq psi1-lowest}
\partial_t \Psi_1 = - \partial_\theta \frac{\hat{\bm{z}}\cdot\bm{\Gamma}^{(0)}_{\text{int}}}{\xi_{\text{r}}} + D_{\text{r}}\, \partial_\theta^2 \Psi_1 + \mathcal{O}(\bm{\nabla}).
\end{equation}
Here, for simplicity, we have dropped the subindex that was labeling particle number 1. We do this hereafter.

\subsection{Hydrodynamic fields and polarity growth rate}

The results obtained above allow us to calculate the growth rate $a$ of the polarity field $\bm{P}$, namely the coefficient of the lowest-order term in the hydrodynamic equation for $\bm{P}$ (\cref{eq polarity}). To this end, we define the hydrodynamic fields, which are the angular moments of the one-particle distribution $\Psi_1$. The two lowest-order moments correspond to the density and the polarity fields:
\begin{subequations} \label{eq hydrodynamic-fields}
\begin{align}
\rho(\bm{r},t) &= \int \Psi_1 (\bm{r},\theta,t) \,\dd\theta,\\
\bm{P}(\bm{r},t) &= \int \hat{\bm{n}}(\theta)\, \Psi_1 (\bm{r},\theta,t)\, \dd\theta.
\end{align}
\end{subequations}
In two dimensions, given that $\hat{\bm{n}}(\theta) = (\cos\theta,\sin\theta)^T$, these fields can be expressed in terms of the angular Fourier components of $\Psi_1$. Based on the Fourier expansion in \cref{eq Fourier-expansion}, the components are given by
\begin{equation} \label{eq angular-Fourier-components}
\tilde\Psi_{1,k} (\bm{r},t) = \int_0^{2\pi} \dd\theta \, e^{ik\theta} \,\Psi_1(\bm{r},\theta,t).
\end{equation}
Thus, the density and polarity fields are given by the zeroth and first components as
\begin{subequations} \label{eq fields-components}
\begin{align}
\rho(\bm{r},t) &= \tilde\Psi_{1,0}(\bm{r},t),\\
\bm{P}(\bm{r},t) &= \left( \begin{array}{c} \Re \tilde\Psi_{1,1}(\bm{r},t) \\ \Im \tilde\Psi_{1,1}(\bm{r},t)\end{array}\right).
\end{align}
\end{subequations}

To obtain an equation for the angular Fourier moments $\Psi_{1,k}(\bm{r},t)$, we project \cref{eq psi1-lowest} according to the integral in \cref{eq angular-Fourier-components}. Using integration by parts, we obtain
\begin{equation}
\partial_t \tilde\Psi_{1,k} = \frac{ik}{2\pi \xi_{\text{r}}} \sum_{m=0}^\infty \tilde\Psi_{1,m} \,\tau_{0,m}\, \tilde\Psi_{1,k-m} - D_{\text{r}} k^2 \tilde\Psi_{1,k},
\end{equation}
where we have used \cref{eq collective-torque-lowest-coefficients}, and with $\tau_{0,m}$ given in \cref{eq torque-coefficients}. The equation for the polarity field $\bm{P}$ then follows from the $k=1$ component. Among the terms in the sum over $m$, we keep only the lowest-order terms $m=0$ and $m=1$. Given that $\tau_{0,0} = 0$, we arrive at
\begin{equation}
\partial_t \tilde\Psi_{1,1} = \frac{i}{2\pi \xi_{\text{r}}} \tilde\Psi_{1,1} \,\tau_{0,1}\, \tilde\Psi_{1,0} - D_{\text{r}} \tilde\Psi_{1,1}.
\end{equation}
Hence, using \cref{eq fields-components}, the equation for the polarity reads
\begin{equation}
\partial_t \bm{P} = \frac{\rho}{2\pi \xi_{\text{r}}} \tau_0 \bm{P} - D_{\text{r}} \bm{P} + \mathcal{O}(\bm{\nabla}),
\end{equation}
where we have defined $\tau_0 \equiv i \,\tau_{0,1}$, whose explicit expression is given in \cref{eq tau0}. Comparing to \cref{eq polarity}, this result provides the growth rate of the polarity field, which is given by \cref{eq growth-rate} as a functional of the density field $\rho$.

\section{Boltzmann kinetic theory} \label{Boltzmann}

In this section, we provide details of the Boltzmann kinetic theory that we use to predict the onset of flocking from binary scattering events (\cref{Fig 5}). We follow Ref. \cite{Lam2015} and we summarize the key steps of the derivation here for reference. We start from the Boltzmann equation for the evolution of the one-particle orientation distribution $f(\theta_1, t)$, where $\theta_1$ is the particle orientation angle. This distribution relates to the full one-particle distribution $\Psi_1 (\bm{r}_1,\theta_1, t)$ in \cref{derivation} as $f(\theta_1, t) = \int \dd^2 \bm{r}_1\, \Psi_1(\bm{r}_1,\theta_1,t)$. The orientation distribution evolves as
\begin{equation} \label{eq Boltzmann}
\partial_t f(\theta_1, t) = I_{\text{scatt}} \left[ f,f\right] + I_{\text{diff}} \left[ f\right],
\end{equation}
where $I_{\text{scatt}} \left[ f,f\right]$ and $I_{\text{diff}} \left[ f\right]$ are functionals representing the changes in particle orientation due to binary scattering events and rotational diffusion, respectively.

We then use \cref{eq Boltzmann} to derive the dynamics of the global polarity $\bm{P}(t) = \int \hat{\bm{n}}_1 (\theta_1)\, f(\theta_1, t) \, \dd\theta_1$, with $\hat{\bm{n}}_1$ the particle orientation vector \cite{Lam2015}. A scattering event changes the polarity by an amount $\delta\bm{p}$, which depends on the geometry of the scattering event parametrized by the incoming angle difference $\theta = \theta_2 - \theta_1$ and half-angle $\bar\theta = \arg\left( e^{i\theta_1} + e^{i\theta_2}\right)$ (\cref{Fig scattering}). Respectively, in a rotational diffusion event, a particle changes its orientation by a random angle $\eta$ with probability $\mathcal{P}_\eta(\eta)$, which produces a polarity change $\delta\bm{p}_{\text{diff}}(\theta_1,\eta) = \mathcal{R}_\eta \hat{\bm{n}}_1 (\theta_1) - \hat{\bm{n}}_1 (\theta_1)$, with $\mathcal{R}_\eta$ being the corresponding rotation matrix. Then, the equation for the polarity reads
\begin{equation} \label{eq Boltzmann-polarity}
\frac{\dd \bm{P}}{\dd t} = \gamma\, \Phi^{\text{scatt}}_f \left[ \delta\bm{p} (\bar\theta, \theta) \right] + \gamma_{\text{diff}}\, \Phi_f^{\text{diff}} \left[ \delta \bm{p}_{\text{diff}} (\theta_1, \eta) \right],
\end{equation}
where $\gamma$ and $\gamma_{\text{diff}}$ are characteristic rates of the scattering and diffusion processes, and the corresponding functionals are given by
\begin{subequations} \label{eq Boltzmann-functionals}
\begin{align}
\Phi_f^{\text{scatt}} \left[\ldots\right] & = \int_0^{2\pi} \dd \bar\theta \int_0^{2\pi} \dd \theta\, K(\theta) f(\theta_1,t) f(\theta_2,t) \left(\ldots\right), \label{eq scattering-functional}\\
\Phi_f^{\text{diff}} \left[\ldots\right] & = \int_0^{2\pi} \dd \theta_1 \int \dd \eta\, \mathcal{P}_\eta(\eta) f(\theta_1,t) \left(\ldots\right). \label{eq diffusion-functional}
\end{align}
\end{subequations}
Here, we took the molecular chaos hypothesis, which assumes that the distributions of two particles before a scattering event are independent. Hence, the two distribution functions $f$ in the scattering functional factorized as seen in \cref{eq scattering-functional}. The additional factor $K(\theta)$ in \cref{eq scattering-functional} is the form factor of a scattering event with angle difference $\theta$. For particles that undergo scattering events at a fixed impact parameter, like in our case, the form factor is $K(\theta) = |\sin (\theta/2)|$, which can be obtained from the Boltzmann cylinder construction \cite{Lam2015,Kardar2007a}.

To obtain an equation for the polar order $P = |\bm{P}|$, we project \cref{eq Boltzmann-polarity} along the polarity direction $\hat{\bm{P}}$, which yields \cite{Lam2015}
\begin{equation} \label{eq Boltzmann-polar-order}
\frac{\dd P}{\dd t} = \rho v_0 r_{\text{int}}\, \Phi_f^{\text{scatt}} \left[ \left(\hat{\bm{p}}\cdot \delta \bm{p} \right) \cos\bar\theta\, \right] - D_{\text{r}} P.
\end{equation}
Here, we used that the characteristic scattering rate $\gamma$ in \cref{eq Boltzmann-polarity} is given by $\gamma = \rho v_0 r_{\text{int}}$ for a gas of self-propelled particles with speed $v_0$, concentration $\rho$, and interaction range $r_{\text{int}}$, which acts as the impact parameter (see \cref{methods} and \cref{Fig scattering-geometry}). The second term in \cref{eq Boltzmann-polar-order} is obtained by explicit integration of the diffusion functional. It describes the loss of polar order, with a coefficient that we identify with the rotational diffusivity $D_{\text{r}}$ of the particles. The explicit integration gives $D_{\text{r}}$ in terms of the rate $\gamma_{\text{diff}}$ in \cref{eq Boltzmann-polarity} and the angular noise distribution $\mathcal{P}_\eta(\eta)$:
\begin{equation} \label{eq rotational-diffusivity}
D_{\text{r}} = \gamma_{\text{diff}} \left( 1 - \int \dd\eta\, \mathcal{P}_\eta(\eta) \cos\eta \right).
\end{equation}

To evaluate the scattering functional, we need to know the orientation distribution $f(\theta_1,t)$. To this end, we take an ansatz of the form $f(\theta_1,t) = f_{P(t)}(\theta_1)$, and we assume that the distribution of orientations in the isotropic phase is uniform up to the constraint
\begin{equation} \label{eq constraint}
\left| \int_0^{2\pi} \hat{\bm{n}}_1(\theta_1) \, f_P(\theta_1) \dd\theta_1 \right| = P.
\end{equation}
The distribution that satisfies these conditions is known as the von Mises distribution, and it is given by \cite{Lam2015}
\begin{equation} \label{eq von-Mises}
f_P(\theta_1) = \frac{ e^{\kappa(\theta_1) \cos\theta_1} }{ 2\pi I_0(\kappa (P))},
\end{equation}
with $\kappa(P)$ determined from
\begin{equation} \label{eq von-Mises-condition}
\frac{I_1(\kappa(P))}{I_0(\kappa(P))} = P,
\end{equation}
where $I_n$ are the modified Bessel functions of the first kind and order $n$.

We then use this distribution in the scattering functional in \cref{eq Boltzmann-polar-order} and expand around the isotropic state to derive the equation for the polar order to linear order. The von Mises distribution expands into
\begin{equation} \label{eq von-Mises-expansion}
f_P(\theta_1) \approx \frac{1}{2\pi} \left( 1 + 2 P \cos\theta_1\right).
\end{equation}
Introducing all the results above into \cref{eq Boltzmann-polar-order}, we obtain
\begin{equation} \label{eq Boltzmann-first-order}
\frac{\dd P}{\dd t} \approx a P,
\end{equation}
with
\begin{multline} \label{eq Boltzmann-growth-rate}
a = \frac{\rho v_0 r_{\text{int}}}{(2\pi)^2} \int_0^{2\pi} \dd \bar\theta \int_0^{2\pi} \dd \theta \left|\sin(\theta/2)\right| \,\bm{p}\cdot\delta\bm{p} (\bar\theta,\theta) - D_{\text{r}} \\
 = \frac{\rho v_0 r_{\text{int}}}{2\pi} \int_{-\pi}^\pi \dd \theta \left| \sin\left(\theta/2\right)\right| \left\langle \bm{p}\cdot \delta\bm{p}(\bar\theta,\theta)\right\rangle_{\bar\theta} - D_{\text{r}}.
\end{multline}
The first term is the change in polar order due to scattering events as provided in \cref{eq momentum-change} in the Main Text.

\onecolumngrid

\clearpage

\setcounter{equation}{0}
\setcounter{figure}{0}
\renewcommand{\theequation}{S\arabic{equation}}
\renewcommand{\thefigure}{S\arabic{figure}}

\onecolumngrid
\begin{center}
\textbf{\large Supplementary Materials}
\end{center}

\section*{Supplementary Figures} \label{SI figures}

\begin{figure*}[!htb]
\begin{center}
\includegraphics[width=0.9\textwidth]{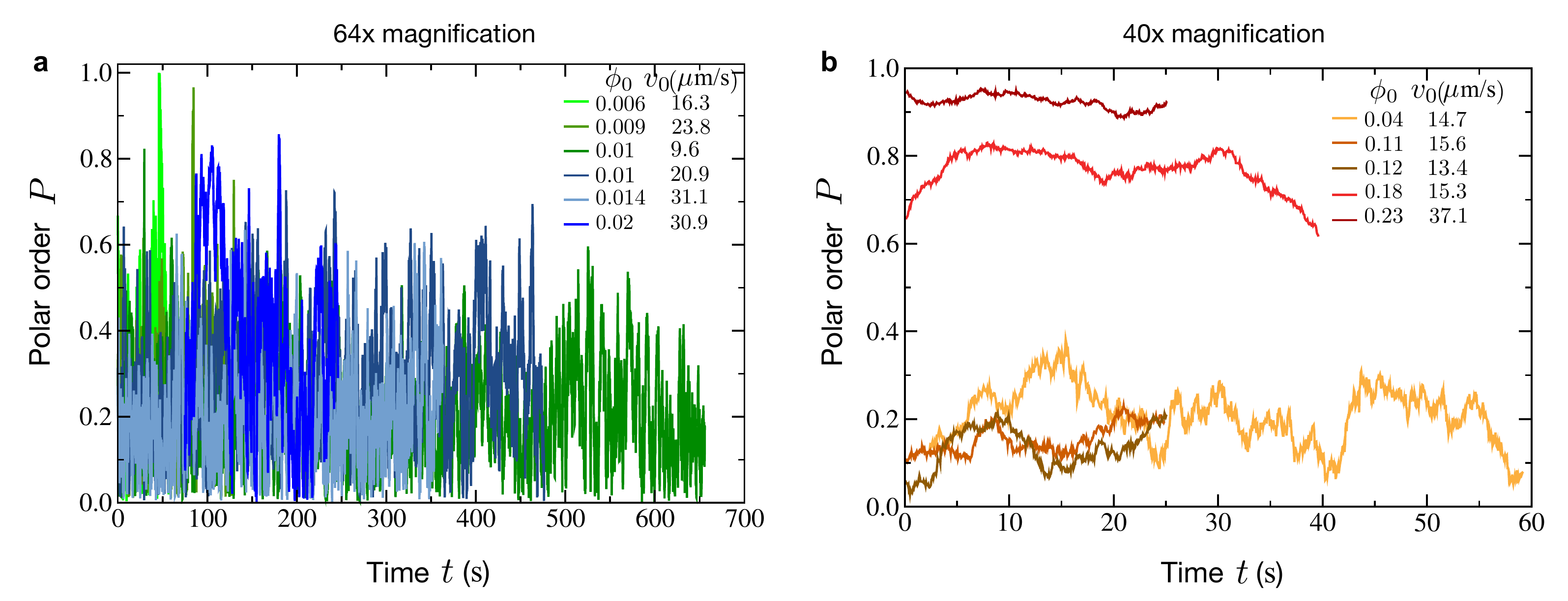}
  {\phantomsubcaption\label{Fig polar-order-64x}}
  {\phantomsubcaption\label{Fig polar-order-40x}}
\end{center}
\bfcaption{Polar order in high-magnification movies}{ Time traces of the polar order parameter $P(t)$ measured from different experimental realizations recorded at 64$\times$ magnification (\subref*{Fig polar-order-64x}) and at 40$\times$ magnification (\subref*{Fig polar-order-40x}). The strong fluctuations in panel \subref*{Fig polar-order-64x} are due to the low number of particles seen in those movies.} \label{Fig experiment-polar-order}
\end{figure*}

\begin{figure*}[!htb]
\begin{center}
\includegraphics[width=0.9\textwidth]{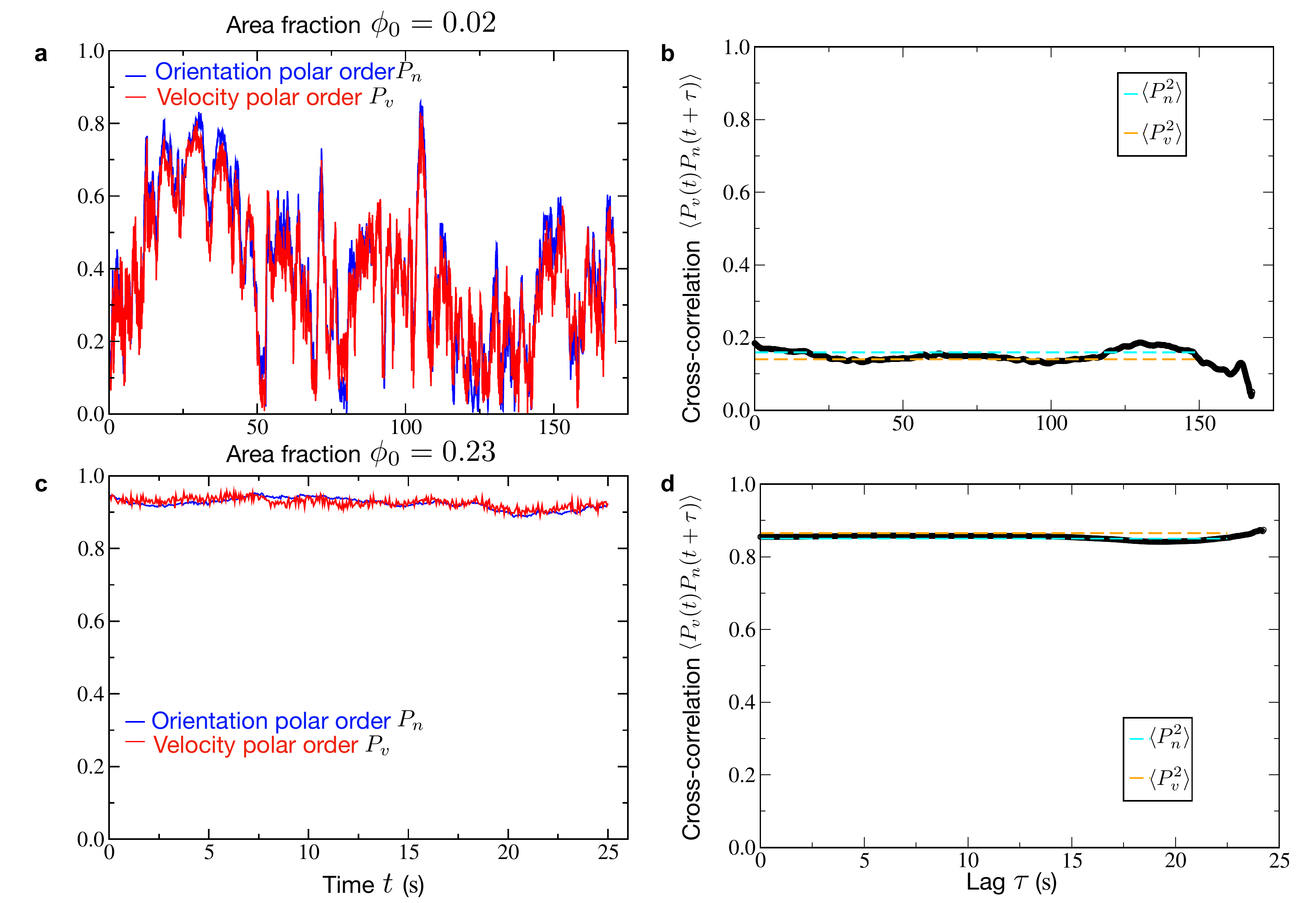}
\end{center}
\bfcaption{Polar order parameters measured from particle orientation and from particle velocity correlate strongly}{ Comparison of time traces of the polar order parameter $P_n$ obtained from particle orientations $\hat{\bm{n}}_i(t)$ (blue) and the polar order parameter $P_v$ obtained from the particle velocity axis $\hat{\bm{v}}_i(t)$ (red). At both high and low area fractions, both order parameters correlate strongly. Accordingly, their cross-correlation stays close to the average of either squared polar order (dashed lines). Because of such high correlation, we can study the emergence of flocking, i.e. polar order in the particle orientations, via the velocity field at low magnifications, which do not allow us to identify particle orientations $\hat{\bm{n}}_i(t)$.} \label{Fig velocity-polarity-correlation}
\end{figure*}

\begin{figure*}[!htb]
\begin{center}
\includegraphics[width=0.9\textwidth]{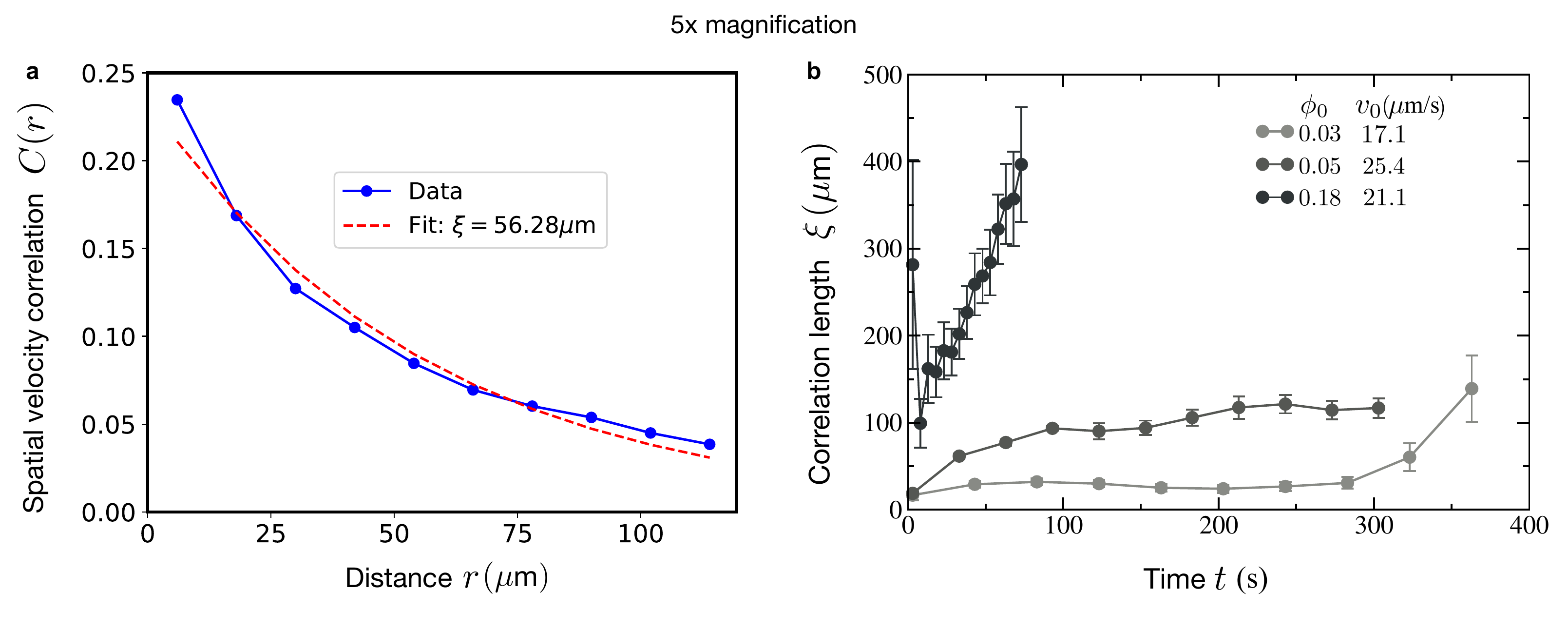}
  {\phantomsubcaption\label{Fig correlation-length-fit}}
  {\phantomsubcaption\label{Fig examples-correlation-length}}
\end{center}
\bfcaption{Velocity correlation length in low-magnification movies}{ \subref*{Fig correlation-length-fit}, Example of an exponential fit to the velocity correlation function $C(r) = \left\langle \bm{v}(r)\cdot \bm{v}(0) \right\rangle$ measured using PIV from low-magnification (5$\times$) experimental movies. This example has values of $\phi_0=0.05$ and $v_0 = 20.4$ $\mu$m/s. From these fits, we obtain the velocity correlation length $\xi(t)$ as a function of time. \subref*{Fig examples-correlation-length}, Three examples of time traces of the correlation length obtained from the fits. At lower area fraction (light gray), the correlation length remains relatively low, corresponding to the isotropic state characterized by disordered particle motion. At higher area fraction (dark gray), the correlation length is much higher and it increases over time, corresponding to the flocking state in which ordered domains coarsen over time.} \label{Fig correlation-length}
\end{figure*}

\begin{figure*}[!htb]
\begin{center}
\includegraphics[width=\textwidth]{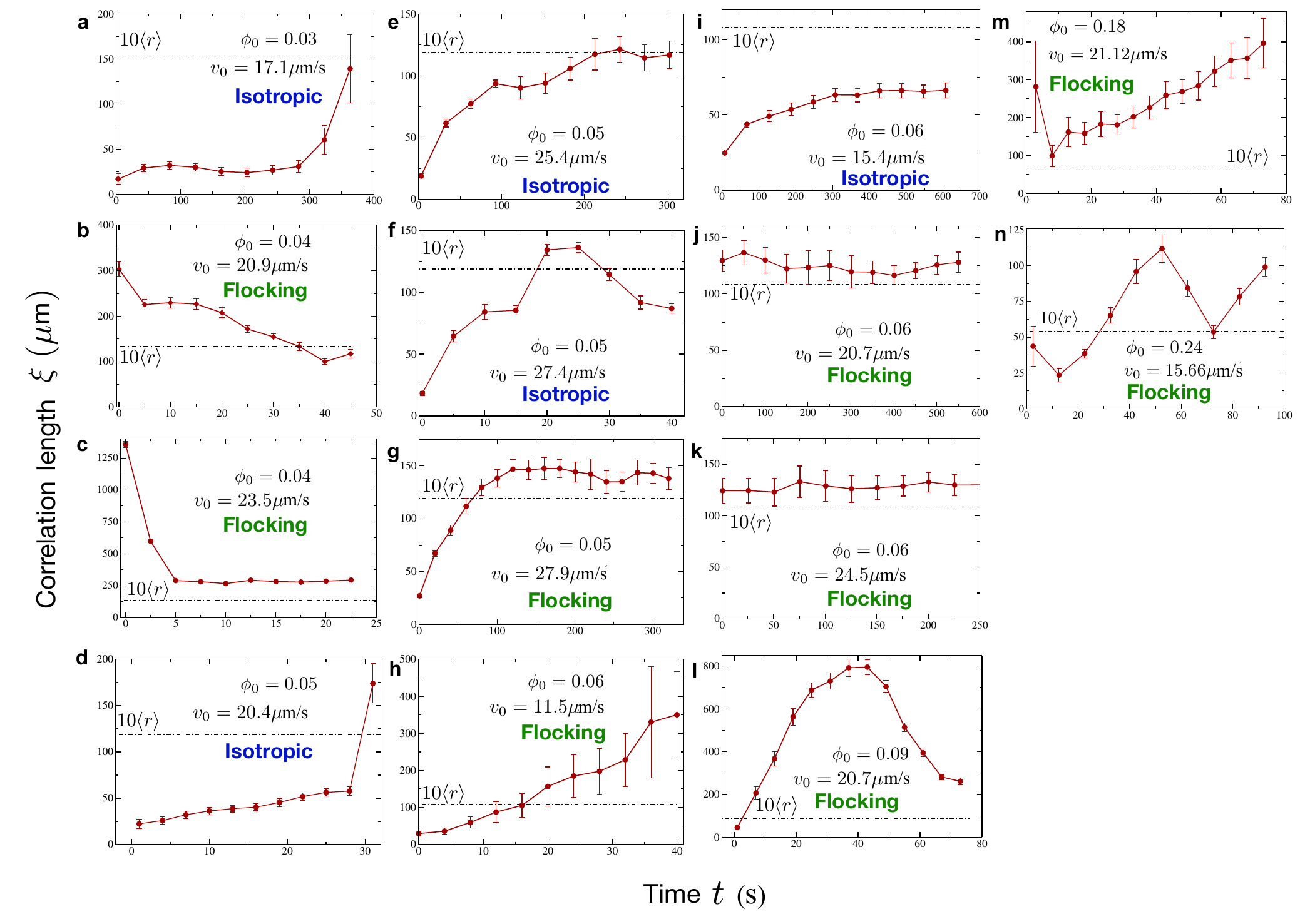}
\end{center}
\bfcaption{Full set of velocity correlation length traces obtained from low-magnification movies}{ We classify each experimental realization imaged at low magnification as being in the flocking state if its velocity correlation length is most of the time larger than $10$ times the average interparticle distance $\left\langle r\right\rangle = \sqrt{1/\rho} = R\sqrt{\pi/\phi_0}$ at each particular area fraction.} \label{Fig correlation-length-analysis}
\end{figure*}

\begin{figure*}[!htb]
\begin{center}
\includegraphics[width=0.9\textwidth]{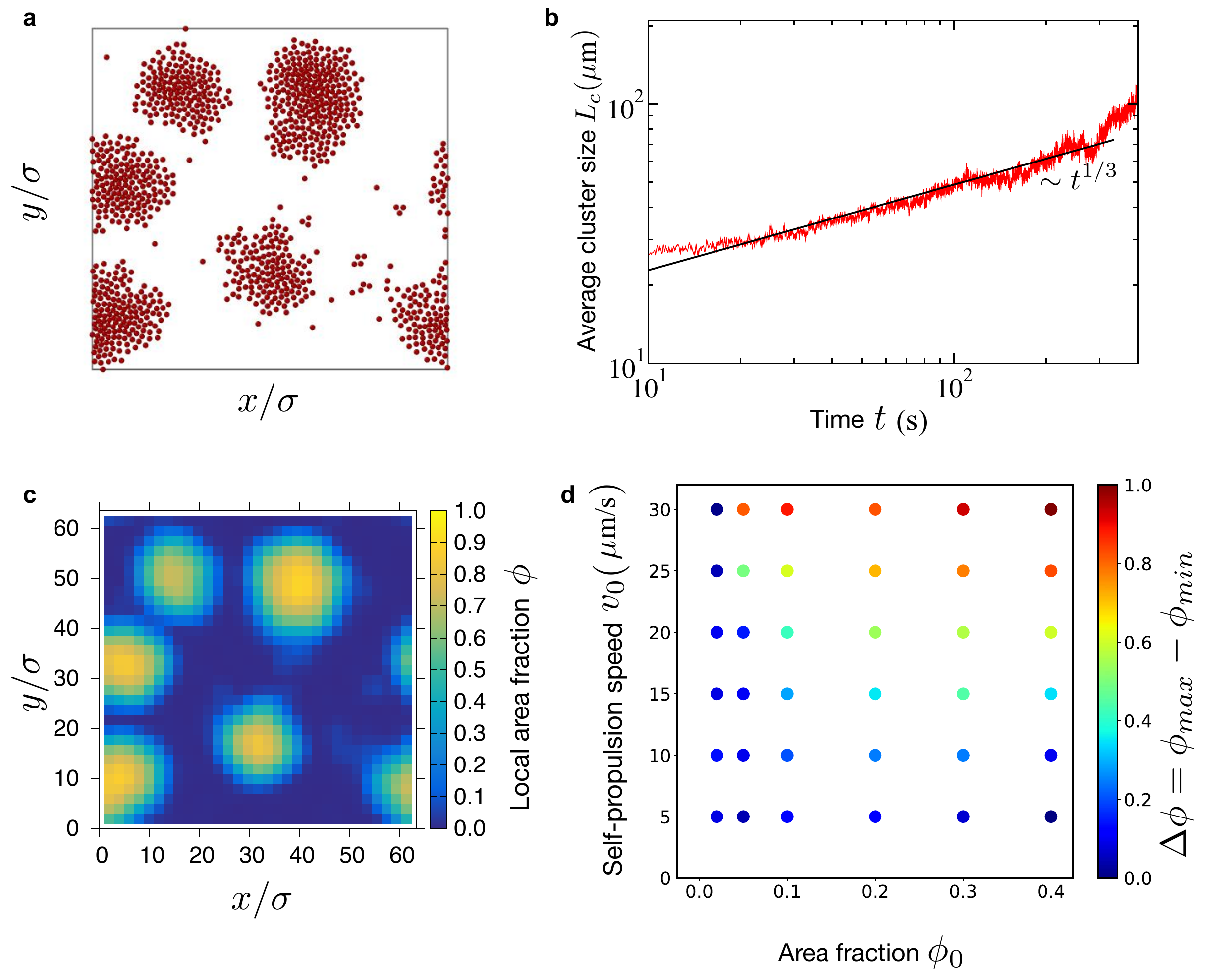}
  {\phantomsubcaption\label{Fig clusters}}
  {\phantomsubcaption\label{Fig coarsening}}
  {\phantomsubcaption\label{Fig area-fraction-field}}
  {\phantomsubcaption\label{Fig phase-separation}}
\end{center}
\bfcaption{Active phase separation of particles that turn towards each other}{ \subref*{Fig clusters}, Simulation snapshot showing clusters formed by particles that turn towards each other, as in Ref.\cite{Zhang2021i}. In this case, the particles follow the same dynamics as in \cref{eq Langevin-main} with \cref{eq force,eq torque} in the Main Text but switching the sign of the interaction torques in \cref{eq torque}. This simulation is for area fraction $\phi_0 = 0.2$ and self-propulsion speed $v_0 = 30$ $\mu$m/s. Other parameter values are listed in \cref{t parameters}. \subref*{Fig coarsening}, Coarsening of the clusters, whose average size grows following Lifshitz-Slyozov scaling law $L_{\text{c}}(t) \sim t^{1/3}$. \subref*{Fig area-fraction-field}, Time-averaged local area fraction, which shows the dense and dilute phases. \subref*{Fig phase-separation}, State diagram of active phase separation based on turn-towards torques. The color code indicates the difference in maxima and minima of local area fraction: $\Delta \phi = \phi_{\text{max}} - \phi_{\text{min}}$.} \label{Fig turn-towards}
\end{figure*}

\begin{figure*}[!htb]
\begin{center}
\includegraphics[width=0.6\textwidth]{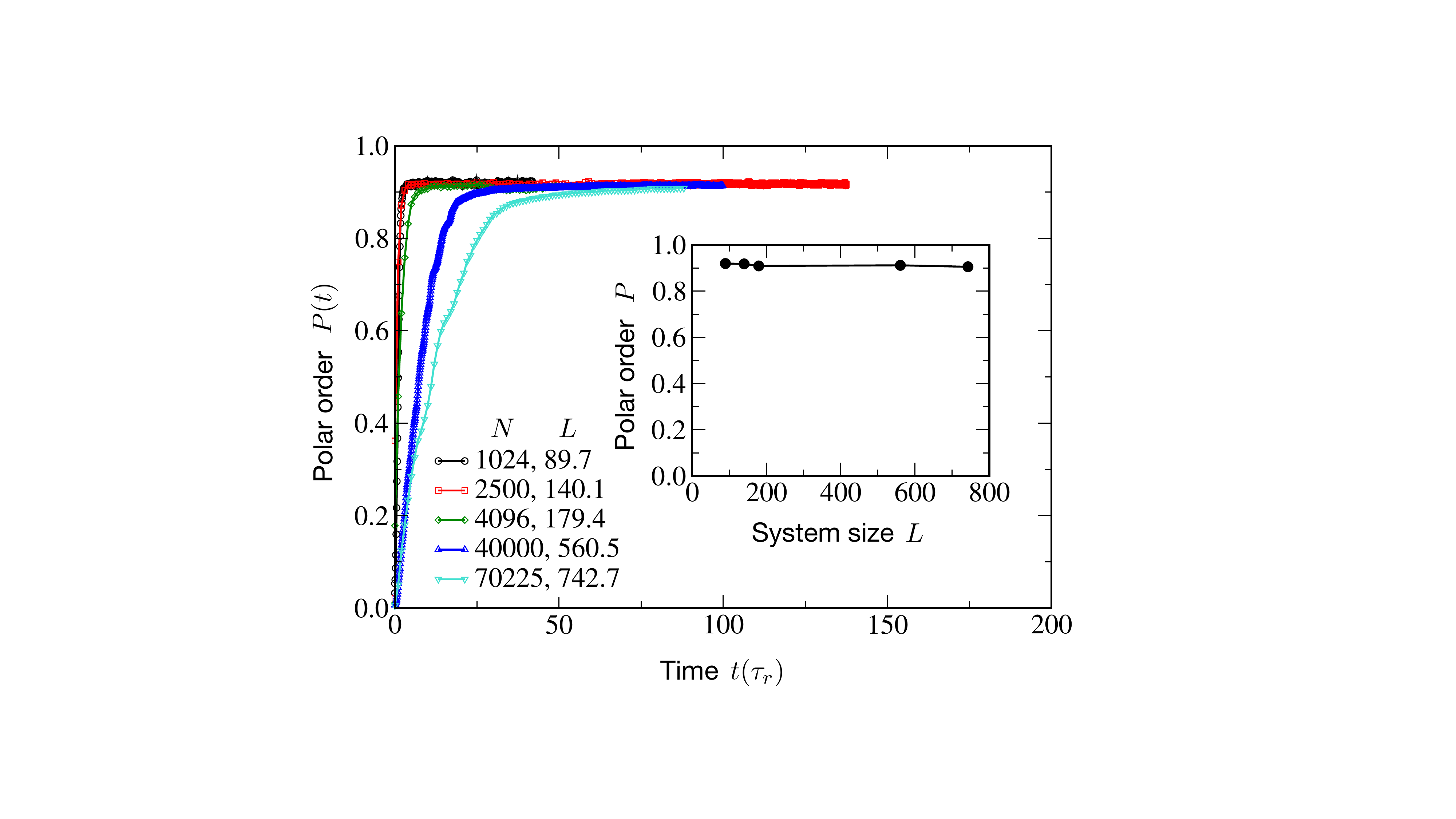}
\end{center}
\bfcaption{Polar order for large system sizes}{ The polar order parameter reaches the same value for simulations of increasing number of particles $N$, corresponding to increasing system size $L$. The simulations are for area fraction $\phi_0=0.1$ and self-propulsion speed $v_0=60$ $\mu$m/s.} \label{Fig system-size}
\end{figure*}

\begin{figure*}[!htb]
\begin{center}
\includegraphics[width=0.5\textwidth]{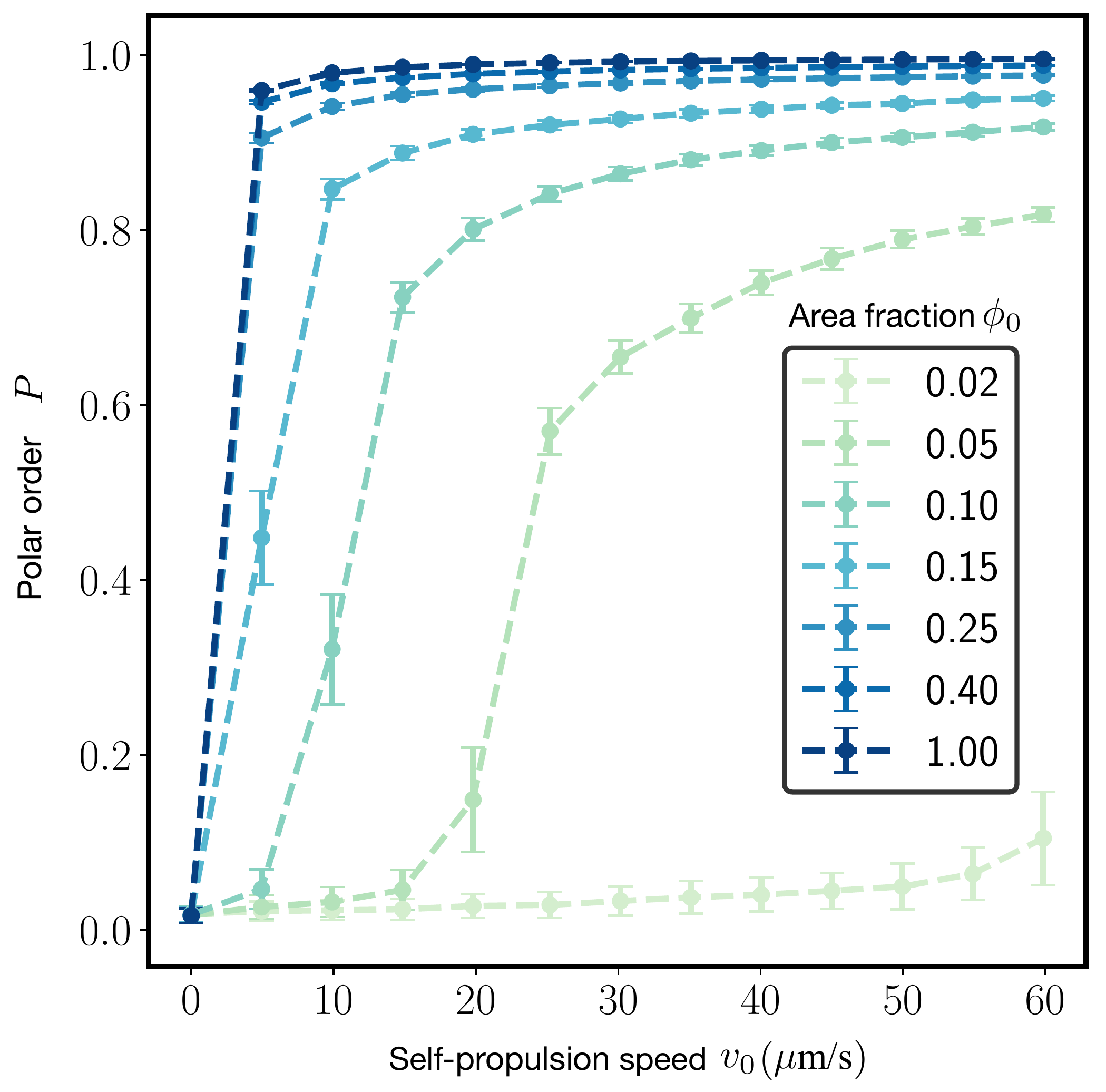}
\end{center}
\bfcaption{Flocking transition}{ Polar order parameter measured in simulations for different self-propulsion speeds and area fractions. For each area fraction, we identify the flocking transition as the steepest point of these curves.} \label{Fig flocking-transition}
\end{figure*}

\begin{figure*}[!htb]
\begin{center}
\includegraphics[width=\textwidth]{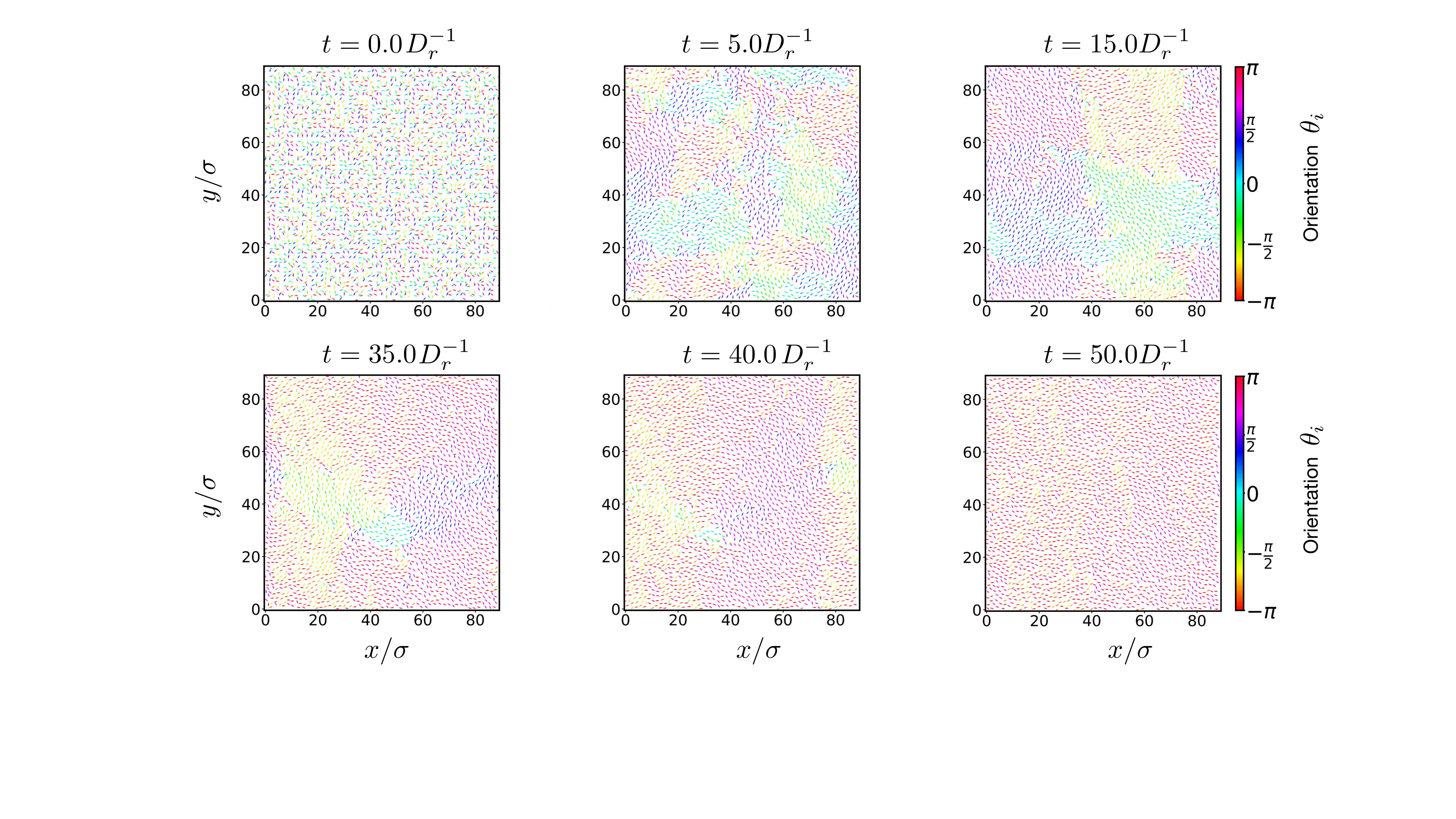}
\end{center}
\bfcaption{Emergence of flocking}{ Series of simulation snapshots showing the emergence of flocking. Polar domains form and coarsen into a uniform flock. The simulation is for $N=2500$ particles at self-propulsion speed $v_0 = 5$ $\mu$m/s.} \label{Fig emergence-flocking}
\end{figure*}

\begin{figure*}[!htb]
\begin{center}
\includegraphics[width=\textwidth]{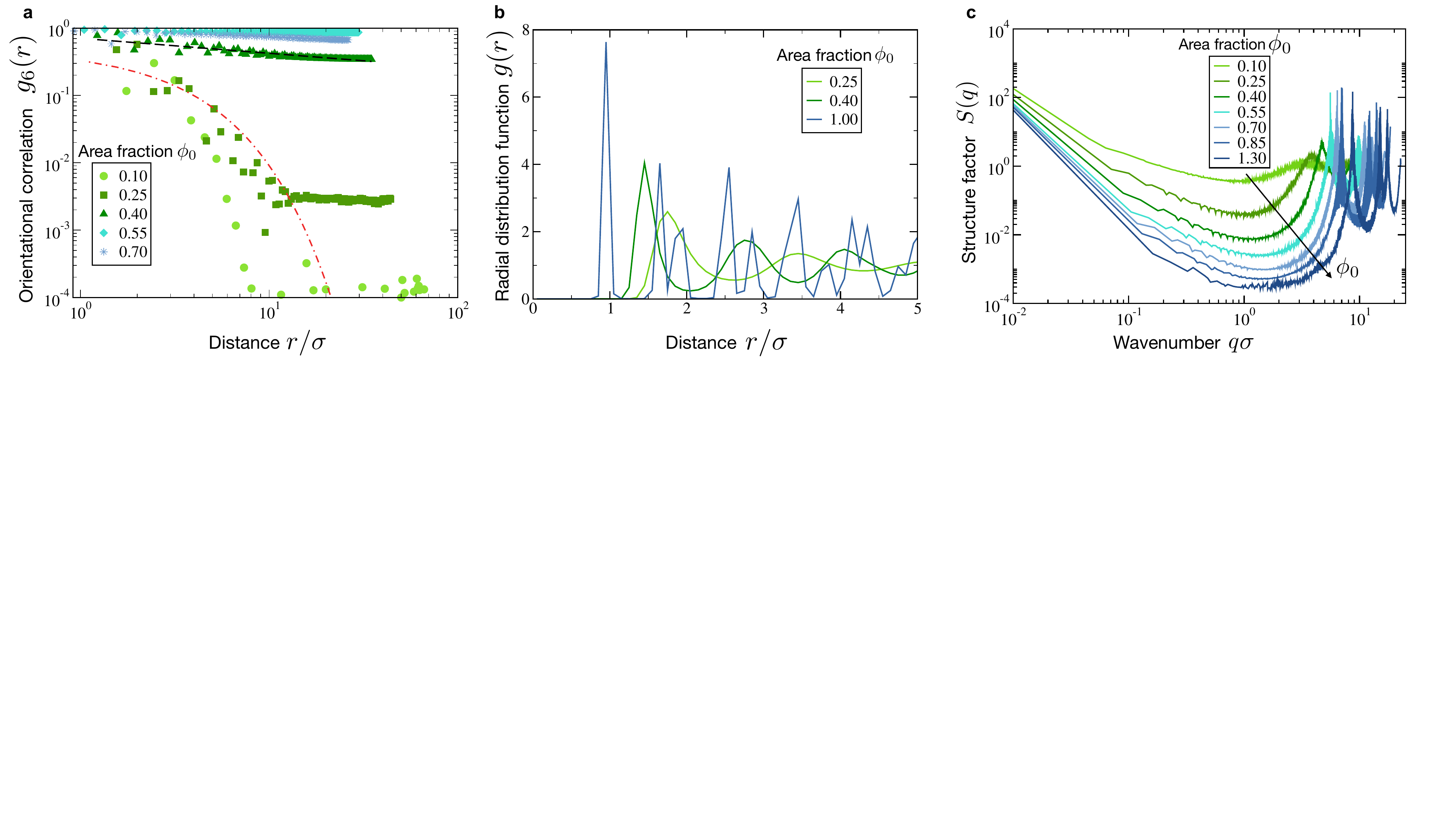}
  {\phantomsubcaption\label{Fig orientational-correlation}}
  {\phantomsubcaption\label{Fig gr}}
  {\phantomsubcaption\label{Fig structure-factor}}
\end{center}
\bfcaption{Orientational and positional correlations}{ \subref*{Fig orientational-correlation}, The orientational correlation function $g_6(r) = \left\langle \psi_6 (\bm{r}') \psi^*_6 (\bm{r}'+\bm{r})\right\rangle/\left\langle \psi_6 (\bm{r}') \psi^*_6 (\bm{r}')\right\rangle$, with $r=|\bm{r}|$, decays exponentially in the fluid phase at low area fractions but as a power law in the hexatic phase at higher area fractions. The red and black dashed curves are exponential and power-law fits, respectively. \subref*{Fig gr}, The radial distribution function develops sharp peaks when the system crystallizes at high area fractions (blue). In panels \subref*{Fig orientational-correlation} and \subref*{Fig gr}, the self-propulsion speed $v_0 = 20$ $\mu$m/s. \subref*{Fig structure-factor}, The structure factor develops peaks at short distances (large $q$) when the system crystallizes at high area fractions (blue). In this panel, $v_0 = 55$ $\mu$m/s.} \label{Fig correlations}
\end{figure*}

\begin{figure*}[!htb]
\begin{center}
\includegraphics[width=\textwidth]{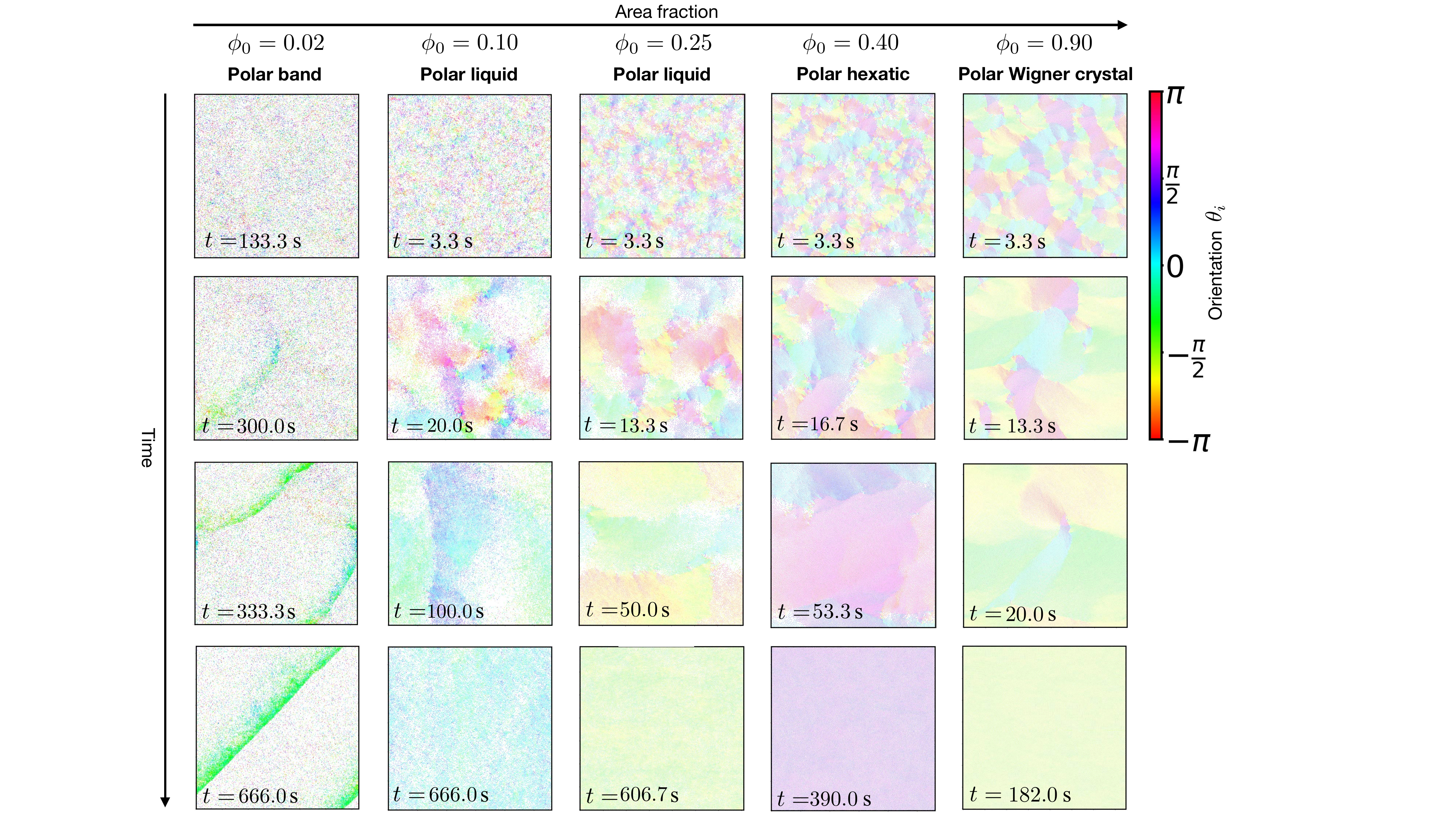}
\end{center}
\bfcaption{Flocks in the form of polar bands, liquids, hexatic states, and active Wigner crystals}{ Series of snapshots showing the emergence of different flocking states in large simulations of $N=40000$ particles at different area fractions. The self-propulsion speed is $v_0=60$ $\mu$m/s. For increasing area fractions, the system flocks in the form of polar bands, as a uniform liquid, as an hexatic state, and as a Wigner crystal, as characterized in \cref{Fig 3}. For increasing area fractions, the evolution towards the final states involves the merging and coarsening of smaller polar bands and vortices, the annihilation of topological defects, and of crystalline grain boundaries.} \label{Fig flocking-states}
\end{figure*}

\begin{figure*}[!htb]
\begin{center}
\includegraphics[width=0.9\textwidth]{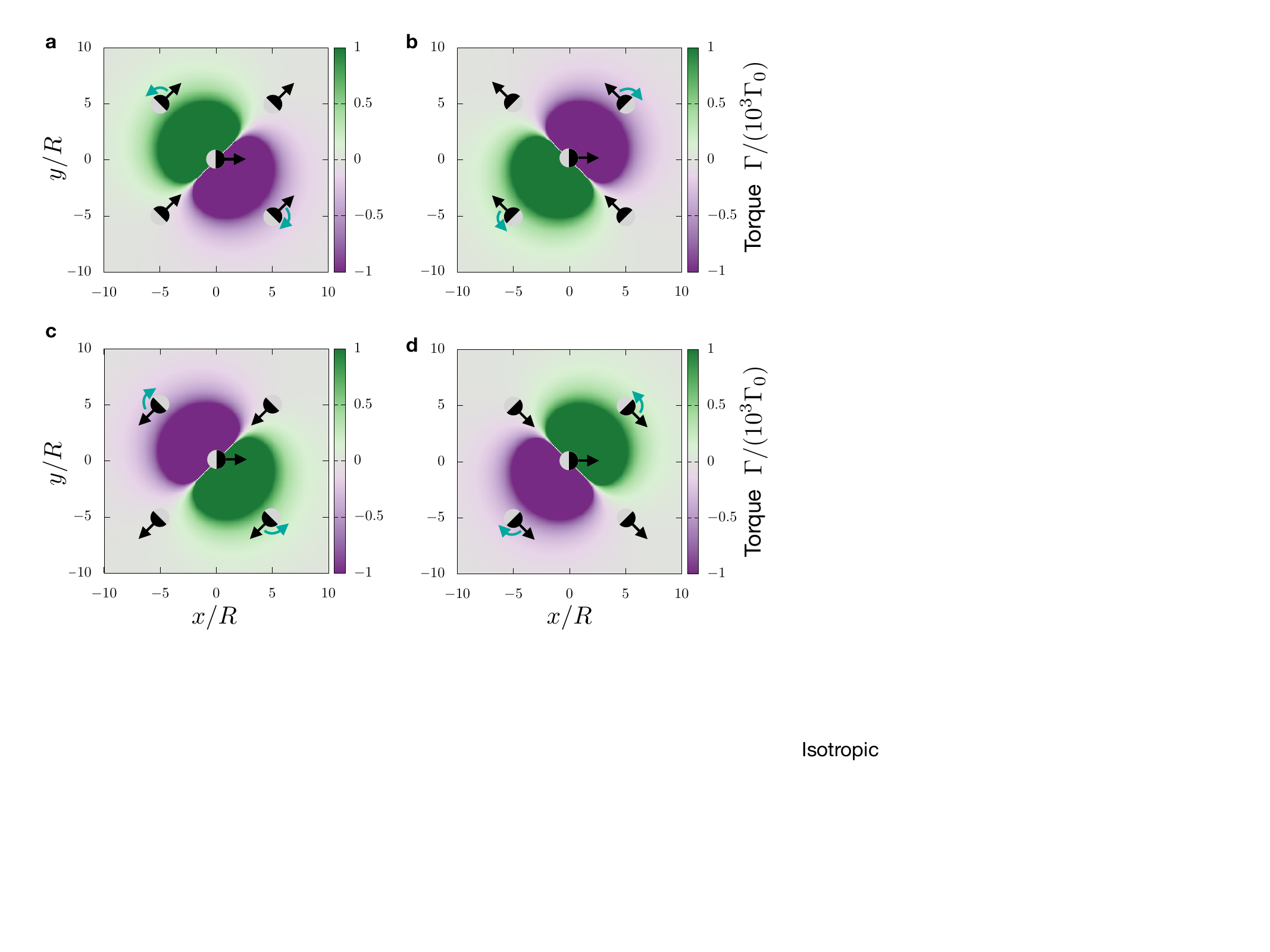}
  {\phantomsubcaption\label{Fig torque-quadrant1}}
  {\phantomsubcaption\label{Fig torque-quadrant2}}
  {\phantomsubcaption\label{Fig torque-quadrant3}}
  {\phantomsubcaption\label{Fig torque-quadrant4}}
\end{center}
\bfcaption{Interparticle torques change with the orientation of the receiving particle}{ \subref*{Fig torque-quadrant1}-\subref*{Fig torque-quadrant4}, Torque fields exerted by the reference particle (center) on a probe particle at position $(x,y)$ with relative orientation $\theta = \pi/4, 3\pi/4$, $-3\pi/4$, and $-\pi/4$, as drawn. These orientations represent each of the four quadrants, which are displayed as those in \crefrange{Fig quadrant-1}{Fig quadrant-4}. Panel \subref*{Fig torque-quadrant1} is a repetition of \cref{Fig torque}, for completeness. Green arrows indicate the directions of the torque in each of the four representative probe particles. The torque is plotted from \cref{eq torque} and normalized by $\Gamma_0 = 3\ell (d_{\text{h}}^2 - d_{\text{t}}^2)/(4\pi\epsilon R^4)$.} \label{Fig torques}
\end{figure*}

\begin{figure*}[!htb]
\begin{center}
\includegraphics[width=0.9\textwidth]{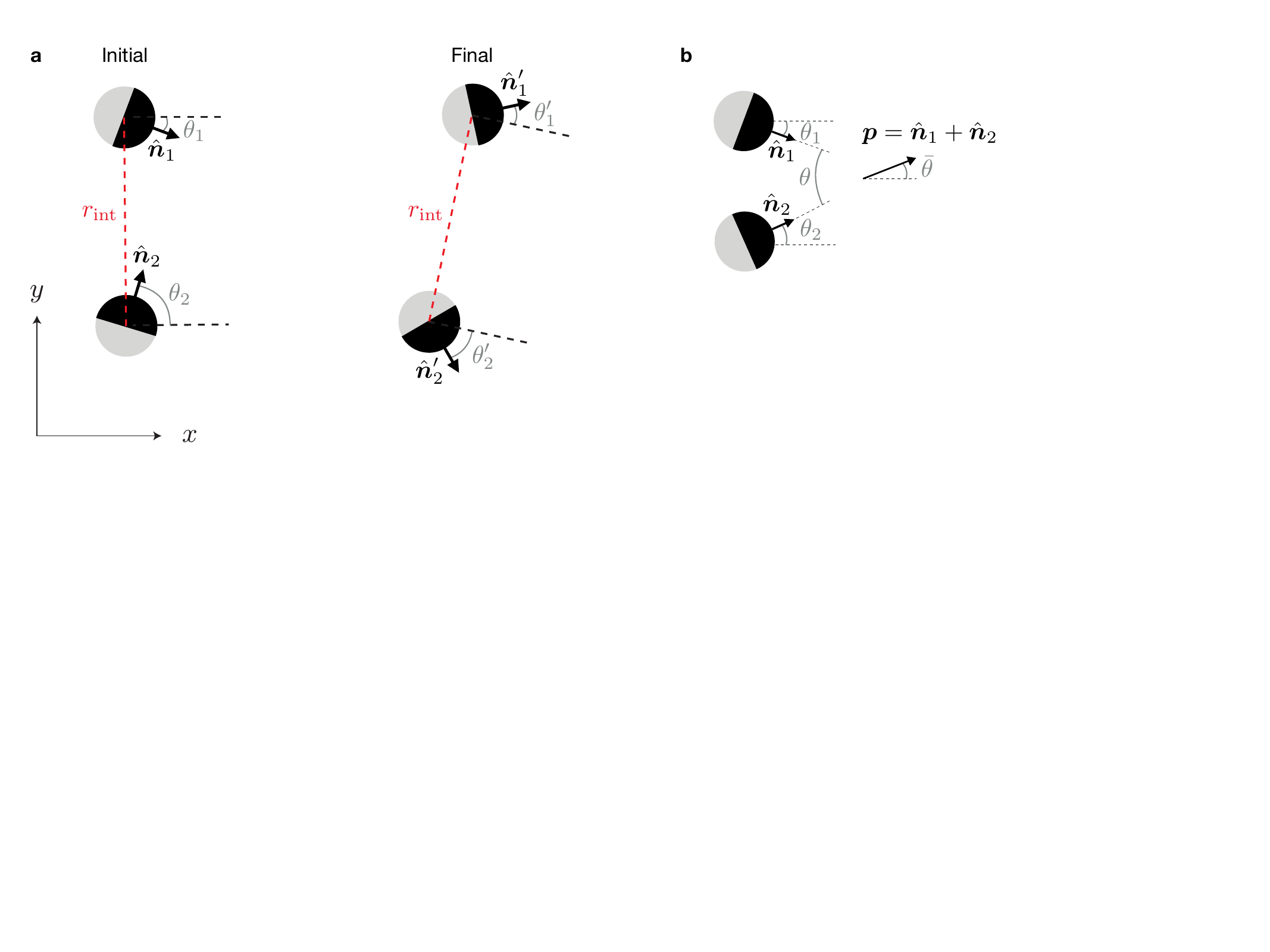}
  {\phantomsubcaption\label{Fig scattering-geometry}}
  {\phantomsubcaption\label{Fig scattering-angles}}
\end{center}
\bfcaption{Scattering geometry}{ \subref*{Fig scattering-geometry}, Initial and final configuration of a pair of particles in a scattering event. \subref*{Fig scattering-angles}, Schematic showing the initial half-angle $\bar\theta = \arg(e^{i\theta_1} + e^{i\theta_2})$, angle difference $\theta = \theta_2 - \theta_1$, and momentum $\bm{p} = \hat{\bm{n}}_1 + \hat{\bm{n}}_2$ of a scattering event.} \label{Fig scattering}
\end{figure*}

\clearpage

\section*{Supplementary Movies} \label{movies}

\noindent\textbf{Movie 1: Isotropic active gas at low area fraction.} Isotropic state imaged with a 64$\times$ objective at 20 frames per second. The particle area fraction is $\phi_0 = 0.01$ and the self-propulsion speed is $v_0 = 20.9$ $\mu$m/s.

\bigskip

\noindent\textbf{Movie 2: Flocking at high area fraction.} Flocking state imaged with a 40$\times$ objective at 20 frames per second. The particle area fraction is $\phi_0 = 0.23$ and the self-propulsion speed $v_0 = 37.1$ $\mu$m/s.

\bigskip

\noindent\textbf{Movie 3: Vortices in the flocking state.} Particles often form vortices, here as part of the flocking state imaged with a 5$\times$ objective at 10 frames per second. The particle area fraction is $\phi_0 = 0.06$ and the self-propulsion speed is $v_0 = 20.7$ $\mu$m/s.

\bigskip

\noindent\textbf{Movie 4: Polar band in the flocking state.} A polar band traveling through the surrounding dilute isotropic gas imaged with a 5$\times$ objective at 10 frames per second. The particle area fraction is $\phi_0 = 0.09$ and the self-propulsion speed is $v_0 = 20.7$ $\mu$m/s.

\bigskip

\noindent\textbf{Movie 5: Isotropic state in simulations.} Simulation showing the isotropic state, shown in \cref{Fig isotropic-simulation}. This simulation is for $N = 2500$ particles at an area fraction $\phi_0 = 0.1$ and self-propulsion speed $v_0 = 5$ $\mu$m/s. The arrows and color code indicate the orientation of particles. Time is in units of $D_{\text{r}}^{-1}$.

\bigskip

\noindent\textbf{Movie 6: Emergence of flocking in simulations.} Simulation showing the emergence of flocking, which involves the formation of polar domains that then coarsen until a they merge into a uniform flock, shown in \cref{Fig flocking-simulation}. This simulation is for $N = 2500$ particles at an area fraction $\phi_0 = 0.25$ and self-propulsion speed $v_0 = 5$ $\mu$m/s. The arrows and color code indicate the orientation of particles. Time is in units of $D_{\text{r}}^{-1}$. Snapshots of this movie are shown in \cref{Fig emergence-flocking}.

\bigskip

\noindent\textbf{Movie 7: Polar bands in simulations.} Simulation showing the emergence of polar bands in a large dilute system with $N = 40000$ particles at an area fraction $\phi_0 = 0.02$ and self-propulsion speed $v_0 = 60$ $\mu$m/s. Time is in units of $D_{\text{r}}^{-1}$. Color indicates the orientation of particles.

\bigskip

\noindent\textbf{Movie 8: Polar liquid via transient polar bands in simulations.} Simulation showing the emergence of multiple polar bands that end up merging into a rather uniform liquid. The simulation is for $N = 40000$ particles at an area fraction $\phi_0 = 0.10$ and self-propulsion speed $v_0 = 60$ $\mu$m/s. Time is in units of $D_{\text{r}}^{-1}$. Color indicates the orientation of particles.

\bigskip

\noindent\textbf{Movie 9: Dense polar liquid in simulations.} Simulation showing the emergence of a dense polar liquid for $N = 40000$ particles at an area fraction $\phi_0 = 0.25$ and self-propulsion speed $v_0 = 60$ $\mu$m/s. Time is in units of $D_{\text{r}}^{-1}$. Color indicates the orientation of particles.

\bigskip

\noindent\textbf{Movie 10: Flocking hexatic state in simulations.} Simulation showing the emergence of a flocking hexatic state for $N = 40000$ particles at an area fraction $\phi_0 = 0.40$ and self-propulsion speed $v_0 = 60$ $\mu$m/s. Time is in units of $D_{\text{r}}^{-1}$. Color indicates the orientation of particles.

\bigskip

\noindent\textbf{Movie 11: Flocking Wigner crystal in simulations.} Simulation showing the emergence of a flocking Wigner crystal. The system initially forms multiple crystalline domains that merge into a single one through the annihilation of grain boundaries. The simulation is for $N = 40000$ particles at an area fraction $\phi_0 = 0.90$ and self-propulsion speed $v_0 = 60$ $\mu$m/s. Time is in units of $D_{\text{r}}^{-1}$. Color indicates the orientation of particles.

\end{document}